\documentclass[a4paper,11pt]{article}
\usepackage[titletoc]{appendix}
\usepackage{mathtools,amssymb}
\usepackage[makeroom]{cancel}
\usepackage{graphicx}
\usepackage{epstopdf}
\usepackage{mathrsfs}
\usepackage{enumitem}
\usepackage{stackrel}
\usepackage{float}
\usepackage{booktabs}
\usepackage[normalem]{ulem}  
\usepackage{longtable}
\usepackage{hyperref}
\usepackage[usenames,dvipsnames,table]{xcolor}
\usepackage{empheq}

\DeclareMathAlphabet{\mathpzc}{OT1}{pzc}{m}{it}
 \allowdisplaybreaks
 \allowbreak

\newcommand{\lsim}{\mathrel{\hbox{\rlap{\lower .55ex
\hbox{$\sim$}} \kern-.3em \raise.4ex \hbox{$<$}}}}
\newcommand{\gsim}{\mathrel{\hbox{\rlap{\lower.55ex
\hbox{$\sim$}} \kern-.3em \raise.4ex \hbox{$<$}}}}

\begin{document}


\newcommand{\partiald}[2]{\dfrac{\partial #1}{\partial #2}}
\newcommand{\be}{\begin{equation}}
\newcommand{\ee}{\end{equation}}
\newcommand{\f}{\frac}
\newcommand{\s}{\sqrt}
\newcommand{\p}{\partial}
\newcommand{\lm}{\mathcal{L}}
\newcommand{\wm}{\mathcal{W}}
\newcommand{\om}{\mathcal{O}_{n}}
\newcommand{\bea}{\begin{eqnarray}}
\newcommand{\eea}{\end{eqnarray}}
\newcommand{\ba}{\begin{align}}
\newcommand{\ea}{\end{align}}
\newcommand{\ep}{\epsilon}

\def\gap#1{\vspace{#1 ex}}
\def\be{\begin{equation}}
\def\ee{\end{equation}}
\def\bal{\begin{array}{l}}
\def\ba#1{\begin{array}{#1}}  
\def\ea{\end{array}}
\def\bea{\begin{eqnarray}}
\def\eea{\end{eqnarray}}
\def\beas{\begin{eqnarray*}}
\def\eeas{\end{eqnarray*}}
\def\del{\partial}
\def\eq#1{(\ref{#1})}
\def\fig#1{Fig \ref{#1}}
\def\re#1{{\bf #1}}
\def\bull{$\bullet$}
\def\nn{\nonumber}
\def\ub{\underbar}
\def\nl{\hfill\break}
\def\ni{\noindent}
\def\bibi{\bibitem}
\def\ket{\rangle}
\def\bra{\langle}
\def\vev#1{\langle #1 \rangle}
\def\mattwo#1#2#3#4{\left(\begin{array}{cc}#1&#2\\#3&#4\end{array}\right)}
\def\tgen#1{T^{#1}}
\def\half{\frac12}
\def\floor#1{{\lfloor #1 \rfloor}}
\def\ceil#1{{\lceil #1 \rceil}}

\def\Tr{{\rm Tr}}

\def\mysec#1{\gap1\ni{\bf #1}\gap1}
\def\mycap#1{\begin{quote}{\footnotesize #1}\end{quote}}

\def\Red#1{{\color{red}#1}}

\def\Om{\Omega}
\def\a{\alpha}
\def\b{\beta}
\def\l{\lambda}
\def\g{\gamma}
\def\eps{\epsilon}
\def\Si{\Sigma}

\def\lan{\langle}
\def\ran{\rangle}

\def\bit{\begin{item}}
\def\eit{\end{item}}
\def\benu{\begin{enumerate}}
\def\eenu{\end{enumerate}}

\def\tr{{\rm tr}}


\parindent=0pt
\parskip = 10pt

\def\a{\alpha}
\def\g{\gamma}
\def\G{\Gamma}
\def\b{\beta}
\def\d{\delta}
\def\D{\Delta}
\def\e{\epsilon}
\def\fb{{\bar f}}
\newcommand{\prop}{G_{\L}}
\newcommand{\propn}{G_{\L'}}
\def\k{\kappa}
\def\r{\rho}
\def\rvec{\vec{\rho}}
\def\l{\lambda}
\def\L{\Lambda}
\def\mO{\mathcal{O}}
\def\mbO{{\mathbb O}}
\def\P{\Phi}
\def\s{\sigma}
\def\t{\theta}
\def\z{\zeta}
\def\n{\eta}
\def\del{\partial}
\def\bt{\bar{\t}}
\def\x{{\rm x}}
\def\scA{\mathcal{A}}
\def\scAb{\mathbb{A}}
\def\md{\mathsf{d}}
\def\mL{\mathcal{L}}
\def\mPz{\mathcal{P}_{\Delta z}}
\def\mQz{\mathcal{Q}_{\Delta z}}
\def\mRz{\mathcal{R}_{\Delta z}}
\def\mK{\mathcal{K}}
\def\Gcoff{\mathbb{G}}
\newcommand{\mD}{\mathcal{D}}
\def\fb{\bar{f}}
\def\gb{\bar{g}}
\def\tl{\tilde{\l}}
\def\pt{\tilde{\phi}}
\def\pb{\bar{\phi}}
\newcommand{\phiO}{\phi^{(0)}}


\newcommand*{\Cdot}[1][1.25]{%
  \mathpalette{\CdotAux{#1}}\cdot%
}
\newdimen\CdotAxis
\newcommand*{\CdotAux}[3]{%
  {%
    \settoheight\CdotAxis{$#2\vcenter{}$}%
    \sbox0{%
      \raisebox\CdotAxis{%
        \scalebox{#1}{%
          \raisebox{-\CdotAxis}{%
            $\mathsurround=0pt #2#3$%
          }%
        }%
      }%
    }%
    \dp0=0pt %
    \sbox2{$#2\bullet$}%
    \ifdim\ht2<\ht0 %
      \ht0=\ht2 %
    \fi
    \sbox2{$\mathsurround=0pt #2#3$}%
    \hbox to \wd2{\hss\usebox{0}\hss}%
  }%
}

\renewcommand{\arraystretch}{2.5}%
\renewcommand{\floatpagefraction}{.8}%
\def\bfnd{\bar{\mathfrak{f}}}
\def\bfd{\mathfrak{f}}
\def\fnd{\bar{f}}
\def\fd{f}
\def\bfndD{\bar{\mathpzc{f}}}
\def\bfdD{\mathpzc{f}}

\def\DCT{\hat{\mathcal{D}}_{ct}}

\def\Jc{\mathcal{J}}
\def\corrM{\mathfrak{M}}
\def\PermP{\mathcal{P}}

\hypersetup{pageanchor=false}
\begin{titlepage}
\begin{flushright}
TIFR/TH/18-16
\end{flushright}

\vspace{.4cm}
\begin{center}
  \noindent{\Large \bf{Quantum quench and thermalization of one-dimensional Fermi gas via phase space hydrodynamics}}\\
\vspace{1cm}
Manas Kulkarni$^a$\footnote{manas.kulkarni@icts.res.in},
Gautam Mandal$^b$\footnote{mandal@theory.tifr.res.in},
and Takeshi Morita$^{c,d}$\footnote{morita.takeshi@shizuoka.ac.jp}

\vspace{.5cm}
\begin{center}
{\it a. International Centre for Theoretical Sciences}\\
{\it Tata Institute of Fundamental Research, Shivakote,
Bengaluru 560089, India.}\\
\vspace{.5cm}
{\it b. Department of Theoretical Physics}\\
{\it Tata Institute of Fundamental Research, Mumbai 400005,
India.}\\
\vspace{.5cm}
{\it c. Department of Physics}\\
{\it Shizuoka University, 836 Ohya, Suruga-ku, Shizuoka 422-8529,
JAPAN.}\\
\vspace{.5cm}
{\it d. Graduate School of Science and Technology}\\
{\it Shizuoka University, 836 Ohya, Suruga-ku, Shizuoka 422-8529,
	JAPAN.}

\end{center}

\gap2

\today

\end{center}

\begin{abstract}

By exploring a phase space hydrodynamics description of one-dimensional free Fermi gas, we discuss 
 how systems settle down to steady states described by the generalized Gibbs ensembles through
quantum quenches. 
We investigate time evolutions of the Fermions which are trapped in external potentials or a circle
 for a variety of initial conditions and quench protocols. 
 We analytically compute local observables such as particle density and show that they always exhibit power law relaxation at late times.  We find a simple rule which determines the power law exponent.
Our findings are, in principle, observable in experiments in an one dimensional
free Fermi gas or Tonk's gas (Bose gas with infinite repulsion).

\end{abstract}

\gap1

\end{titlepage}
\setcounter{page}{1}
\tableofcontents

\section{Introduction}

It is an important issue to find out if and when a closed system,
subjected to some disturbance--- such as in a quantum quench, reaches
equilibrium after some time.  From a theoretical viewpoint, this is a
fundamental question since it lies at the heart of quantum statistical
mechanics; e.g., subtleties arise in the applicability of
thermodynamics in systems showing many-body localization. The question
of thermalization is also important from an experimental viewpoint.
In experiments dealing with ultra-cold atoms, the time scale of
equilibration as well as the nature of the equilibrium are measurable
quantities and help characterize the system in question.  In the
latter context, it has also been realized, for a little over a decade
now, that there is a non-trivial notion of thermalization of local
observables in a so-called free systems or integrable systems where
the thermal ensemble is often characterized by an infinite number of
chemical potentials, corresponding to an infinite number of conserved
quantities. Such an ensemble is called a GGE (generalized Gibbs
ensemble). These issues have been summarized in a number of articles
in recent years; for a partial list of references, see
\cite{Polkovnikov:2010yn, 2016AdPhy..65..239D, 2018arXiv180506422W}.

The issue of thermalization in integrable models \cite{konik1,konik2,konik3,konik4,konik5,konik6, konik7,0295-5075-115-4-40011} has been discussed in
detail some time ago in Ref~\cite{2008PhRvL.100j0601B}, where
thermalization has been shown to occur under some general assumptions.
In spite of such general arguments, it is important to see if one can
obtain some explicit results about time evolution of observables,
especially at long times. For quantum quenches leading to gapless
Hamiltonians, such explicit results were obtained at long times in
Ref. \cite{Mandal:2015jla} which rigorously showed thermalization of
the reduced density matrix of subsystems of a large system. Fully
exact time-dependence and subsequent results on thermalization were
obtained for free scalars and Fermions in 1+1 dimensions for mass
quenches ending at zero mass in \cite{Mandal:2015kxi}.

In this paper, we will discuss quantum quench in a free Fermi gas in
one space dimension, by using a large-$N$ technique\footnote{In some special cases, the thermodynamical properties of free Fermions and free Bosons are similar. (See \cite{doi:10.1119/1.1286116,doi:10.1119/1.1544520} for recent developments.)
	It may be interesting to explore the quantum quenches of free boson systems in which we can observe similar time evolutions to those of the free fermion systems.  
}. This subject was
dealt with earlier in a paper \cite{Mandal:2013id} by two of the
present authors, where it was shown how moments of the Fermion density
approached thermalization from particular sudden quenches. Apart from studies of thermalization, several other
dynamical aspects, such as the onset of shock fronts, have been
studied in large-$N$ non-interacting as well
as interacting Fermi gases 
\cite{kul0,kul1,aag0,aag1,SciPostPhys.2.1.002} (see also the Appendix \ref{sec-collective}) ; finite-$N$ corrections
have been studied in \cite{PhysRevLett.109.260602}.

The importance of the large-$N$ limit is that the Fermions are
described by a semiclassical fluid in the phase space. As it turns
out, for simple phase space configurations, the equations describing
such phase space fluids boil down to equations of conventional
hydrodynamics. Thus, it is tempting to think that thermalization can
be understood somehow in terms of the equations of conventional
hydrodynamics, which would be of significant interest. 
However, conventional hydrodynamics of the Fermions typically breaks down way before we reach asymptotic times because of formation of shock fronts \cite{Mandal:2013id,kul0,kul1,aag0,aag1, PhysRevLett.109.260602}. (See Figure \ref{fig-fold} also.)
We will show that one can easily proceed beyond such times when one sticks to methods of phase
space hydrodynamics, where shock fronts and singularities in real space merely
become folds in phase space which are smooth, continuous
configurations and do not present any singularity. 
Thus, using the phase space formulation, we will derive results on thermalization of large-$N$ Fermion systems at large times.

The time evolutions of the one-dimensional free Fermi gases which are equivalent to the gases of the hardcore bosons (Tonks-Girardeau gas  \cite{PhysRev.50.955,doi:10.1063/1.1703687, paredes2004tonksgirardeau, Kinoshita1125}) are actively being investigated in cold atomic systems and condensed matter physics\footnote{
	 Experimentally, it is more conceivable to prepare infinitely repulsive bosons than to realize pure 1D non-interacting Fermions. Extending and adapting these ideas to higher dimensions is not straightforward. }. If these gases in one dimension (1D) are confined in an external potential or a periodic circle, they evolve to a steady state which is described by GGE.
 \footnote{An exception is the motion in an external harmonic potential \cite{PhysRevLett.94.240404}. In this case, every particle moves by the common periodicity and the dephasing \cite{2008PhRvL.100j0601B} does not occur.}
There, various observables have been investigated in various quench procedures \cite{PhysRevB.88.205131, PhysRevLett.110.245301, 1742-5468-2013-09-P09025, 1742-5468-2014-1-P01009, 1742-5468-2014-11-P11016, 1742-5468-2014-11-P11016, 1742-5468-2014-12-P12012, PhysRevA.96.023611, PhysRevA.97.033609}. 
We will see that our phase space hydrodynamics approach is quite powerful and makes it possible to reveal the general nature of the late time behaviour of the time evolution.
Particularly we show that the local observables exhibit power law relaxation where the exponent is fixed through a simple universal rule.
This rule works for arbitrary quench procedures and external potentials (except the harmonic one in which the relaxation does not occur).
We demonstrate it in several examples which are summarized in Table \ref{table:power}.
We will also find the entropy formula for the late time GGE state.

\begin{table}
\begin{tabular}{|l | l| c|}
	\hline
	\textit{Quench Protocol}  &   \textit{Power Law Exponent}  &  \textit{Section} \\
	\hline
released from a potential $x^{2m}$ to a periodic circle &  $t^{-\frac{2m+1}{2m}}$ &  \ref{sec:release} \\
	\hline
	released from a box potential to a periodic circle &  $t^{-1}$ &  \ref{sec:box} \\
	\hline
	introduction of a cosine potential from $V=0$ &  $t^{-3/2}$ &  \ref{sec-cos} \\
\hline
\end{tabular}
\caption{Examples of  power law relaxation of one point functions, e.g. particle densities, at late
	 times.
The results may work for arbitrary local observables defined by  \eqref{F-local}.
}
\label{table:power}
  \end{table}

It is worth mentioning that the two main ingredients of our setup, namely external potentials and quenches are both experimentally realistic. Potentials and confinements of various kinds such as harmonic trap, quartic traps \cite{quart2,quart0,quart1}, ring-shaped traps \cite{ringPRL,ring1}, box-like traps \cite{almostbox,zh0,Navon167,navon17b,navon17a,navon17}, and sinusoidal potentials have been realized. Various quench protocols have been successfully demonstrated \cite{2012PhRvL.109q5301C,konik_prl_2014,2012PhRvL.109q5301C}. 
Besides, quantum quanches can be applied to the so-called ``shortcut to adiabaticity" in cooling atoms \cite{PhysRevLett.104.063002, 1367-2630-13-11-113017, TORRONTEGUI2013117, Onofrio:2017tlp, PhysRevLett.115.025302, 2018arXiv180701724D}.

The summary and organization of our paper are as follows.  In Section
\ref{sec-large-N}, the phase space hydrodynamics method is introduced in
the limit of large number ($N$) of Fermions. A simple parametric
solution of the phase space density is presented in case where the
confining potential $V$ vanishes.  In Section \ref{sec-V=0}, we
continue the case of $V=0$ for motion of Fermions on a circle.  In
Subsection \ref{sec:relaxation} we find explicit formulae for the
Fermion density and exhibit power law relaxation where the exponent
depends on the initial profile.  In Subsection \ref{sec:relaxation} we
consider free motion after releasing the Fermi gas from a confining
potential of the form $V= x^{2m}$. In this case the Fermion density
approaches equilibrium according a power law determined by $m$. In
Subsection \ref{sec:box}, the Fermions are released from a box, and
the power law is now universal, viz. $\sim 1/t$. In Section
\ref{v-not-zero}, we consider motion of Fermions in a potential at the
post-quench stage. We find in Subsection \ref{power-law-v-nonzero}
that there is power law relaxation even in this case. An explicit example
of quenching from $V=0$ to a cosine potential is shown in Subsection \ref{sec-cos}.
As claimed in the beginning, the post-quench reduced density matrix is expected
to relax to that of a thermal or a GGE state. We explicitly
verify this in Section \ref{sec:GGE} by comparing the
time-evolving phase space density after a long time with that in
a GGE; we also compute, in Subsection \ref{delta-s}, the relevant
entropy production. 
A discussion on conventional hydrodynamics and its breakdown is given in the Appendix \ref{sec-collective}.

\section{Time evolution of free Fermions in a thermodynamical limit}\label{sec-large-N}

In this section, we review the phase space formulation of the dynamics
of a one-dimensional Fermi gas (which is equally applicable to a
hard-core Bose gas in the Tonks limit).

Let us consider $N$ non-relativistic free Fermions in one dimension
whose single particle Hamiltonian is given by
\begin{align}
	\hat h =  -\frac 12 \hbar^2 \partial_x^2 + V(x).
	\label{H-single}
\end{align}
Here $V(x)$ is an external potential.
The physics of this Fermi gas can be described by the second quantized
Fermion field
\begin{align}
	& \psi(x,t) =  \sum_n \hat{c}_n   \chi_n(x) \exp[-i E_n t/\hbar], \quad
	\hat h \varphi_n(x) = E_n \varphi_n(x),
	\label{psi-x-t}\\
	& i\hbar \del_t \psi(x,t) =  -\frac 12 \hbar^2 \partial_x^2 \psi(x,t)
	+ V(x) \psi(x,t),
	\label{dirac}\\
	&\int dx\ \psi^\dagger (x,t) \psi(x,t) = N.
	\label{n-const}
\end{align}
The dynamics of the Fermions can alternatively be expressed in
terms of the Wigner phase space density which is defined by\footnote{Although the Wigner phase space density is called ``density", it could be negative. However, once we take the large $N$ limit \eqref{large-N}, it becomes always non-negative and we do not have this issue.}
\begin{align}
	u(x,p,t)=  \int d\eta\   \psi^\dagger(x+\eta/2,t)
	\psi(x-\eta/2,t)\ \exp[i\eta p/\hbar].
	\label{wigner-app}
\end{align}
The constraint \eq{n-const} translates to the following
constraints in terms of the $u$-variable:
\begin{align}
	u(x,p,t)*u(x,p,t)= u(x,p,t),\qquad
	\int \frac{dx dp}{2\pi \hbar}  u(x,p,t) = N,
	\label{u-constraints}
\end{align}
whereas the equation of motion \eq{dirac} translates to
\begin{align}
	\f\partial{\partial t} u(x,p,t) + \{h(x,p),u(x,p,t)\}_{_{MB}} = 0.
	\label{u-eom-mb}
\end{align}
The derivation of the above equations is straightforward, and is given
in detail in \cite{Dhar:1992rs, Dhar:1992hr, Dhar:1993jc}. Here we
have used the notation
\begin{align}
	f*g (x,p) & \equiv \big[
	\cos {\hbar \over 2} (\partial_p \partial_{x'} - \partial_{p'}
	\partial_x) (f(x,p,t) g(p',x',t))\big]_{p'=p,x'=x},
	\label{star}
	\\
	\{f, g\}_{MB} &=  f*g - g*f.
	\label{MB}
\end{align}
Indeed, it has been shown in \cite{Dhar:1993jc} that the Fermion
path integral can be entirely rewritten in terms of the $u(x,p,t)$
variable, subject to the constraints \eq{u-constraints}.

It is easy to show that observables of the Fermi fluid, such as the
density $\rho(x,t)$ and the fluid velocity can be expressed
in terms of the phase space density
\begin{empheq}[box=\fbox]{align}
	\rho(x,t)  =  \int \f{dp}{2\pi \hbar} u(x,p,t),
	\quad
	v(x,t)  = \f1{\rho(x,t)} \int \f{dp}{2\pi \hbar}\ p\kern2pt u(x,p,t).
	\label{densities}
\end{empheq}

\subsection{Large $N$ limit of Fermi gas and phase space hydrodynamics}\label{sec:phase-hydro}

\begin{figure}
	\begin{center}
		\includegraphics[scale=0.6]{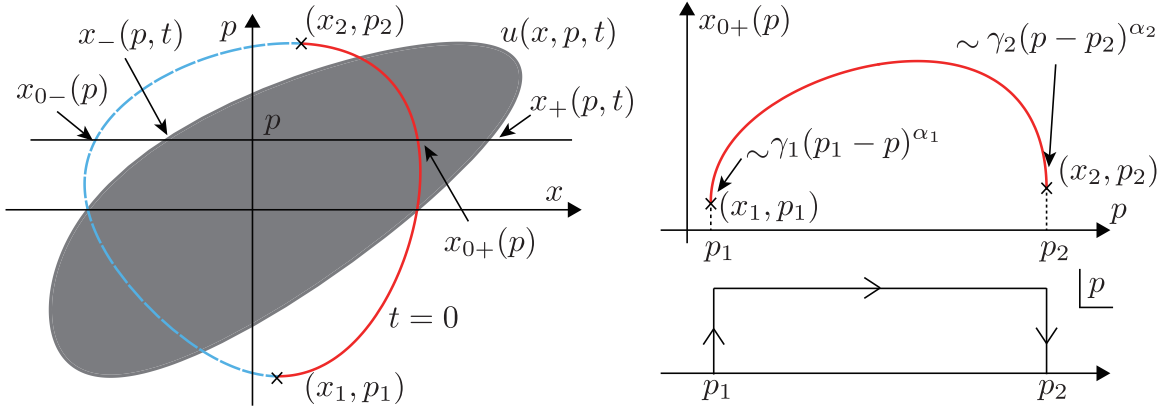}
		\caption{
			(Left) Fermion droplet in the phase space.
			The red-blue curve describes the boundary of the droplet at $t=0$.
			$x_{+}(p,t)$ and $x_{-}(p,t)$ denote the boundaries of the droplet for a given $p$ at time $t$, and 
			$x_{0+}(p)$ (red) and $x_{0-}(p)$ (broken blue line) are those at time $t=0$.
			At $(x_1,p_1)$ and $(x_2,p_2)$, the curves $x_{0+}(p)$ and $x_{0-}(p)$ meet each other.
			(Right Top) Plot of $x_{0+}(p)$.
			(Right Bottom) Integral contour \eqref{integral-p} on the complex $p$-plane. 
		}
		\label{fig:quadratic}
	\end{center}
\end{figure}

We will define the large $N$ limit as
\begin{align}
	N \to \infty, \quad  \hbar \to 0, \quad N\hbar = 1. 
	\label{large-N}
\end{align}
It is easy to see that in this limit the star product in
\eq{star} becomes an ordinary product and the Moyal bracket
\eq{MB} becomes a Poisson bracket. Thus the
phase space density in the large $N$ limit satisfies
the equation of motion 
\begin{align}
	\f\partial{\partial t} u(x,p,t) + \{h(x,p),u(x,p,t)\}_{_{PB}} = 0,
	\label{liouville}
\end{align}
which is simply Liouville's equation for the classical phase space
density.  With $h(x,p)= p^2/2 + V(x)$, this becomes \footnote{If we substitute $u(x,p,t)= 2\pi \hbar \sum_{i=1}^N \delta(x-x_i(t)) \delta(p-p_i(t)) $ into (\ref{u-eom}), we obtain the classical Hamilton equation $\dot x_i =p_i$ and $\dot{p}_i=-V'(x_i)$.  Thus the points inside the droplet correspond to single Fermions.
}
\begin{align}
	\partial_t u + p\partial_x u - V'(x)\partial_pu = 0.
	\label{u-eom}
\end{align}
The constraints \eq{u-constraints} become
\begin{empheq}[box=\fbox]{align}
	u^2= u, \quad  \int \frac{dpdx}{2\pi} \  u(x,p,t)= N\hbar = 1.
	\label{u-constr}
\end{empheq}
The first constraint implies that at any given phase space point
$(x,p)$ the phase space density can be either  =0 or =1. The regions
where $u=1$ are called droplets, representing regions occupied by
Fermions, 1 each in every small cell, of area $\hbar$. $u=0$
represents regions without Fermions (see Figure \ref{fig:quadratic}).
The second constraint implies that the area of a droplet (or in case
of multiple disconnected droplets, combined area of all droplets) is
$N$.  As an example, the Fermi sea for a harmonic trap $V= x^2/2$ is
represented by a circular droplet centred at the origin and of area 1.

\subsection{$u(x,p,t)$ in $V=0$ case} 
In the case of $V=0$, the solution  $u(x,p,t)$ to Eqs. \eqref{u-eom} and \eqref{u-constr} is very simple. 
The solution of the constraint $u^2=u$ is given by using step function
\begin{align}
u(x,p,t)=  \theta(x_+(p,t)- x) \theta(x- x_-(p,t)),
\label{quadratic}
\end{align}
where $x_{\pm}(p,t)$ describe the boundaries of the droplet as  functions of $p$ at time $t$.  See Figure \ref{fig:quadratic}.
If the droplet has multiple boundaries for a given $p$, for example see Figure \ref{fig:stripe}, the corresponding step functions should be added.
Then Eq. \eqref{u-eom} is satisfied if 
\begin{align}
x_{\pm}(p,t)=x_{0\pm}(p)+pt.
\label{sol-free}
\end{align}
Here $x_{0\pm}(p) \equiv x_{\pm}(p,0)$ are the boundaries of the droplet at $t=0$.
This is because the Fermions move obeying the classical equations $\dot x=p$ and $\dot p =0$.
Thus the Fermions with $p>0$ move towards the right and those with $p<0$ move towards the left as sketched in Figure \ref{fig:quadratic}.

We need to choose this initial profile satisfying the second constraint of (\ref{u-constr})
\begin{align}
\int \frac{ dxdp}{2\pi}\  u(x,p,t=0)=\int \frac{ dp}{2\pi}\left( x_{0+}(p)-x_{0-}(p) \right)=1.
\label{ini-constr}
\end{align}
Once we impose this constraint at $t=0$, Liouville's theorem ensures that $u(x,p,t)$ satisfies the constraint for any $t$.

We can solve Eqs. (\ref{u-eom}) and (\ref{u-constr}) in the $V\neq 0$ case similarly, but the solution is a bit complicated. Hence we first consider the time evolution problem in the $V=0$ case and argue the $V\neq 0$ case later.

\section{Time evolution of particles on a circle in the $V=0$ case}
\label{sec-V=0}

\begin{figure}
	\begin{center}
		\includegraphics[scale=0.6]{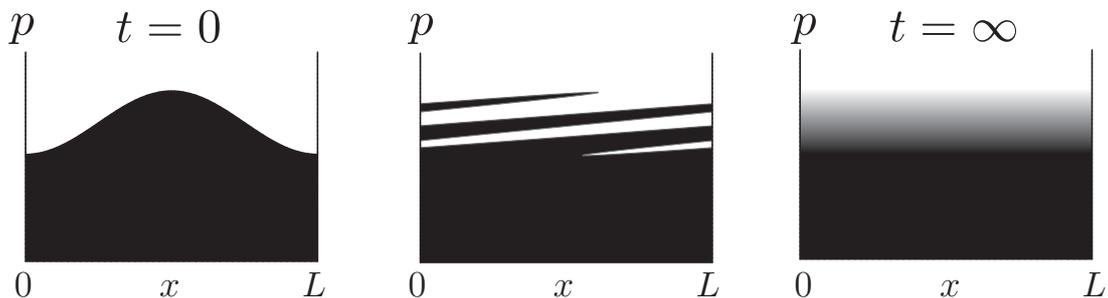}
		\caption{Cartoon of the time evolution of a droplet on a circle in the $V=0$ case.
			The speed of the particle increases as $p$ increases and the droplet will be tilted as time evolves.
			Finally it will be smeared uniformly on the circle and reach a steady state.
		}
		\label{fig:thermalization}
	\end{center}
\end{figure}

We consider the time evolution of the particles in the $V=0$ case.
To make the Fermions confined, we put the Fermions on a circle with a period $L$ and investigate how the particles settle down to a steady state (if it exists). 
By regarding the particle motion \eqref{sol-free}, we can speculate that the system would reach a steady state as shown in Figure \ref{fig:thermalization}.
We will confirm this picture through detailed computations.

Due to the periodicity, we need to modify the previous result (\ref{quadratic}) as
\begin{align}
u(x,p,t)=& \sum_{m} \theta\left(x_+(p,t)- (x+mL)\right) \theta\left(x+mL- x_-(p,t)\right) \nonumber \\
=&\sum_{m} \theta\left(x_{0+}(p)+pt- x-mL\right) \theta\left(x+mL- x_{0-}(p)-pt\right) ,
\label{periodic-u}
\end{align}
where the summation $m$ represents the contributions of the mirrors.
Now we apply Poisson summation formula 
\begin{align}
\sum_{m}f(mL)= \sum_k \frac{1}{L}\int dz f(z) e^{-i \frac{2\pi k z}{L}},
\label{Poisson}
\end{align}
to this equation and obtain
\begin{align}
u(x,p,t)
=&\sum_{k} \frac{1}{L} \int_{-\infty}^\infty dz ~ \theta\left(x_{0+}(p)+pt- x-z\right) \theta\left(x+z- x_{0-}(p)-pt \right) e^{-i\frac{2\pi kz}{L}} \nonumber \\
=&\sum_{k} \frac{1}{L} \int_{x_{0-}(p)+pt- x}^{x_{0+}(p)+pt- x} dz ~ e^{-i\frac{2\pi kz}{L}} \nonumber \\
=& \frac{x_{0+}(p)-x_{0-}(p)}{L}
+   \sum_{k\neq 0} \frac{1}{2 \pi i k}  \left[   
  \exp\left( i \frac{2\pi  k}{L} z \right)   \right]_{z=x_{0-}(p)+pt-x}^{z=x_{0+}(p)+pt-x}.
  \label{u-resum}
\end{align}
This result is suggestive.
While the second terms depend on time $t$, the first term does not.
Particularly, for large $t$, the second terms are highly oscillating and may be irrelevant.
Thus the first term may describe the late time steady state and the second terms may represent the damping terms\footnote{Due to the high oscillations, the width of the stripe  of the Wigner phase space function, e.g., the second panel of Figure \ref{fig-evolution}, may reach the order of $1/N$ and  the finite $N$ corrections may become relevant \cite{PhysRevLett.109.260602, PhysRevA.97.063614}. In order to suppress these corrections, we need to take the large $N$ limit first and, after that, take the large $t$ limit to see the thermalization. Hence the order of these two limits does not commute.}. 

To see it more concretely, we apply this formula to the particle density (\ref{densities})  and obtain
\begin{align}
\rho(x,t) 
 =& \int \frac{dp}{2\pi \hbar} u(x,p,t) \nonumber \\
=& \frac{N}{L}
+  \sum_{k\neq 0} \int \frac{ dp}{4 \pi^2 i \hbar k}     
   \left( \exp\left( i \frac{2\pi k}{L}\left( x_{0+}(p)+pt-x\right) \right)
   -\exp\left(i \frac{2\pi k}{L}\left( x_{0-}(p)+pt-x\right) \right) \right) ,
  \label{rho-resum}
\end{align}
where we have used \eqref{ini-constr}.
Since the second terms would be suppressed at large $t$ through the $p$ integral  due to the oscillation, we would obtain
\begin{align}
&\rho(x,t) 
= \frac{N}{L}, \qquad (t \to \infty).
\label{rho-steady}
\end{align}
This result indicates that the $N$ particles uniformly spread over the circle independent of the initial profile, and this is exactly what we have expected in Figure \ref{fig:thermalization}.
Thus it supports our conjecture that the first term in the formula (\ref{u-resum}) describes the late time steady state. 

\subsection{Power law relaxation}
\label{sec:relaxation}

Now we evaluate the time dependent terms of $\rho(x,t)$ (\ref{rho-resum}) at large $t$ to see whether they are really suppressed and, if so, how they damp as the system evolves.
We will show that it exhibits a power law relaxation and the exponent is fixed by the extrema of the initial droplet at $t=0$.
In the case of the droplet shown in Figure \ref{fig:quadratic}, the behaviors around the minimum $x_1$  and maximum $x_2$ are relevant. 
We investigate this case as an example.

Suppose that  $x_{0+}(p)$ can be expanded around $x_1$ and $x_2$ as 
\begin{align}
	x_{0+}(p) & = x_1 + \gamma_1 (p-p_1)^{\alpha_1} + \cdots , \qquad (p\sim p_1), \nonumber \\
	x_{0+}(p) & = x_2+ \gamma_2 (p_2-p)^{\alpha_2} + \cdots,  \qquad (p\sim p_2),
	\label{profile-edge}
\end{align}
where $\gamma_i$ and $\alpha_i$ ($i=1,2$) are positive constants.
If the droplet is smooth, $0<\alpha_i<1$ are satisfied.
See Figure \ref{fig:quadratic} (Right).
We evaluate the integral of the second term of \eqref{rho-resum} at large $t$, and consider the third term later.
We calculate the positive $k$ and negative $k$ cases separately.
First we consider the positive $k$. 
Since $x_{0+}(p)$ is defined in $p_1 \le p \le p_2$, the integral becomes
\begin{align}
	&	  \sum_{k=1}^{\infty} \int_{p_1}^{p_2} \frac{ dp}{4 \pi^2 i \hbar k}     
	\exp\left( i \frac{2\pi k}{L}\left( x_{0+}(p)+pt-x\right) \right) \nonumber \\
	=
	&  \sum_{k=1}^{\infty} \frac{1}{4 \pi^2 i \hbar k} \left[  \int_{p_1}^{p_1+i \infty} dp 
	\exp\left( i \frac{2\pi k}{L}\left( x_{0+}(p)+pt-x\right) \right) 
         - 
	\int_{p_2}^{p_2+i \infty} dp     
	\exp\left( i \frac{2\pi k}{L}\left( x_{0+}(p)+pt-x\right) \right)  \right] \nonumber \\
	=
	&  \sum_{k=1}^{\infty} \frac{1}{4 \pi^2  \hbar k} e^{ i \frac{2\pi k}{L}\left( x_1+p_1 t-x\right)}   \int_{0}^{\infty} d\eta    e^{ - \frac{2\pi k }{L} t \eta }
	\left(1+ i \frac{2\pi k}{L} \gamma_1 (i \eta)^{\alpha_1} + \cdots \right) \nonumber \\
	&-  \sum_{k=1}^{\infty} \frac{1}{4 \pi^2  \hbar k} e^{ i \frac{2\pi k}{L}\left( x_2+p_2 t-x\right)}   \int_{0}^{\infty} d\zeta    e^{ - \frac{2\pi k }{L} t \zeta }
	\left(1+ i \frac{2\pi k}{L} \gamma_2 (-i \zeta)^{\alpha_2} + \cdots \right).
	\label{integral-p}
\end{align}
In the second line, we have changed the integral contour as shown in Figure \ref{fig:quadratic} (Right Bottom).
Since $t$ is large, the contributions near the real axis would dominate and we ignore the integral along the horizontal line\footnote{Depending on $x_{0+}(p)$, the integrand of \eqref{integral-p} may diverge when $p \to i \infty$. Besides there may be poles or cuts in the Im$(p) >0$ region.	
	However, for sufficiently large $t$, we can avoid them by taking the horizontal line of the integral contour in a finite region, since only the $\eta, \zeta  \lesssim 1/t$ region contributes to the integral.}.
In the third line, we have defined the variables $i\eta=p-p_1$ and  $i\zeta=p-p_2$.
We have also used \eqref{profile-edge} and assumed that $\eta$ and $\zeta$ are small, since the integral in the large $\eta$ and $\zeta$ regions are exponentially suppressed.
Note that the leading term of the expansion in the integral is canceled by the same term coming from the integral \eqref{rho-resum} concerning $x_{0-}(p)$.
Hence the second term is the leading contribution.
We can perform the integral of the second term by using the gamma function $\Gamma(z)=\int_0^{\infty} dy \ e^{-y} y^{z-1}$, and obtain
\begin{align}
		&  \frac{1}{2 \pi  \hbar L} \sum_{k=1}^{\infty} \left( \gamma_1  \Gamma(\alpha_1+1) \left( \frac{iL}{2\pi kt} \right)^{\alpha_1+1}  e^{ i \frac{2\pi k}{L}\left( x_1+p_1 t-x\right)}  
		+  \gamma_2  \Gamma(\alpha_2+1) \left( \frac{-iL}{2\pi kt} \right)^{\alpha_2+1}  e^{ i \frac{2\pi k}{L}\left( x_2+p_2 t-x\right)}  \right) \nonumber \\
		=& \frac{\gamma_1  \Gamma(\alpha_1+1)}{2 \pi  \hbar L}   \left( \frac{iL}{2\pi t} \right)^{\alpha_1+1} {\rm Li}_{\alpha_1+1} \left( e^{ i \frac{2\pi }{L}\left( x_1+p_1 t-x\right)}   \right) 
		 + \frac{ \gamma_2  \Gamma(\alpha_2+1)}{2 \pi  \hbar L}  \left( \frac{-iL}{2\pi t} \right)^{\alpha_2+1} {\rm Li}_{\alpha_2+1} \left( e^{ i \frac{2\pi }{L}\left( x_2+p_2 t-x\right)}   \right) ,
		\label{rho-polylog}
\end{align}
where we have used the polylogarithm function Li${}_s (z)=\sum_{k=1}^{\infty} z^k/k^s $.

For the negative $k$ in \eqref{rho-resum}, we change the integral contour similar to Figure \ref{fig:quadratic} but Im$(p)<0$ region, and obtain
\begin{align*}
&  
\frac{\gamma_1 \Gamma(\alpha_1+1)}{2 \pi  \hbar L}   \left( \frac{-iL}{2\pi t} \right)^{\alpha_1+1} {\rm Li}_{\alpha_1+1} \left( e^{- i \frac{2\pi }{L}\left( x_1+p_1 t-x\right)}   \right)
 + \frac{\gamma_2  \Gamma(\alpha_2+1) }{2 \pi  \hbar L} \left( \frac{iL}{2\pi t} \right)^{\alpha_2+1} {\rm Li}_{\alpha_2+1} \left( e^{ -i \frac{2\pi }{L}\left( x_2+p_2 t-x\right)}   \right)  .
\end{align*}
By adding them to \eqref{rho-polylog}, we finally obtain.
\begin{align}
	\frac{\gamma_1 }{2 \pi  \hbar L}  \left( \frac{L}{ t} \right)^{\alpha_1+1} \zeta\left( -\alpha_1,  \frac{x-x_1-p_1t}{L} -\left\lfloor \frac{x-x_1-p_1t}{L} \right\rfloor \right) \nonumber \\
	+
	\frac{\gamma_2 }{2 \pi  \hbar L}  \left( \frac{L}{ t} \right)^{\alpha_2+1} \zeta\left( -\alpha_2,  \frac{x_2+p_2t-x}{L} -\left\lfloor \frac{x_2+p_2t-x}{L} \right\rfloor \right),
	\label{rho-x+}
\end{align}
where we have used Hurwitz's formula
\begin{align}
i^{-s} {\rm Li}_{s}\left( e^{2\pi i x} \right) + i^s {\rm Li}_{s}\left( e^{-2\pi i x} \right) = \frac{(2 \pi)^s  }{\Gamma(s)} \zeta (1-s,x-\lfloor x \rfloor),
\end{align}
where $\zeta(s,z)$ is the Hurwitz zeta function and $\lfloor x \rfloor$ is the floor function.
Note that $x- \lfloor x \rfloor$ is a periodic function\footnote{
	$x- \lfloor x \rfloor$ is singular at $x=n$  $(n \in {\mathbf Z})$.
	Correspondingly Eq. \eqref{rho-x+} has cusp singularities at $x-x_1-p_1 t=nL$ and $p_2t-x+x_2=nL$. 
	These are the location of the peaks of the droplet and the singularities correspond to the shock fronts  at large $t$. See Figure \ref{fig-fold}.}with a period 1.
This is the leading contribution to $\rho(x,t)$ from the integral of the $x_{0+}(p)$ terms in \eqref{rho-resum} at large $t$ and is suppressed by the power law as we announced.

Similarly we have the contributions from $x_{0-}(p)$. If $x_{0-}(p)$ can be expanded near the extrema as
\begin{align}
	x_{0-}(p) & = x_1 - \gamma^{-}_1 (p-p_1)^{\alpha^{-}_1} + \cdots , \qquad (p\sim p_1), \nonumber \\
	x_{0-}(p) & = x_2- \gamma^{-}_2 (p_2-p)^{\alpha^{-}_2} + \cdots , \qquad (p\sim p_2),
\end{align}
where $\gamma^-_i$ and $\alpha^-_i$ ($i=1,2$) are positive constants\footnote{If the shape of the droplet is $C^{\infty}$, $\gamma_i=\gamma_i^-$ and $\alpha_i=\alpha_i^-=1/2m_i$ where $m_i$ are positive integers.}, we obtain the same expression to \eqref{rho-x+} but $\gamma_i \to \gamma_i^-$ and $\alpha_i \to \alpha_i^-$.

Thus the density $\rho(x,t)$ behaves as
\begin{align}
	\rho(x,t) = \frac{N}{L}+O\left(t^{-(\alpha+1)}\right), \qquad \alpha={\rm min}(\alpha_1,\alpha_2,\alpha^-_1,\alpha^-_2).
	\label{density-decay}
\end{align}
Note that a smaller $\alpha_i$ means a flatter extremum of the initial droplet with respect to $x$.
Therefore  exponent of late time power law relaxation is fixed by the flattest extremum.

\begin{figure}
	\begin{center}
		\includegraphics[scale=1.2]{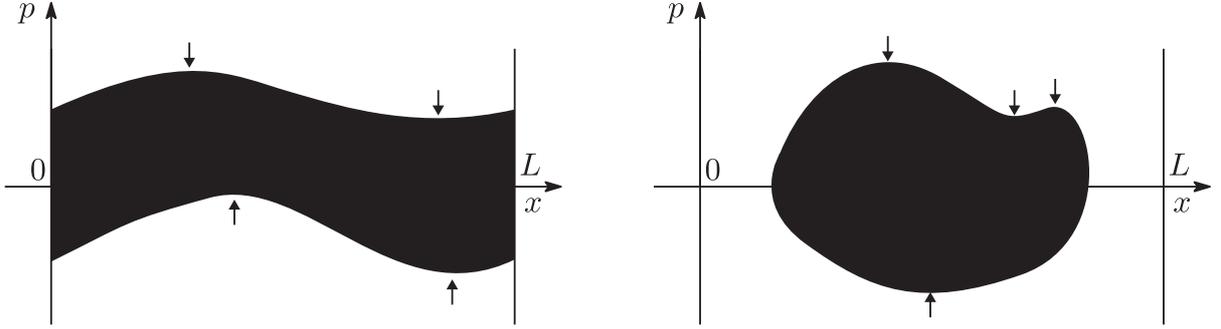}
		\caption{ Fermion droplets at $t=0$ and the singular points in $x_{0\pm}( p)$ at which $\partial_p x_{0\pm}( p)$ diverges.
		}
		\label{fig:stripe}
	\end{center}
\end{figure}
  
If the initial droplet has more than two extrema as shown in Figure \ref{fig:stripe}, 
we should divide the boundary at each extrema and define the boundary function $x(p)$ between them. (Hence we need four boundary functions in the cases of Figure \ref{fig:stripe}.)
Then the calculations are almost the same as the $x_+(p)$ case and obtain a similar result to \eqref{rho-x+} \footnote{If the extremum at $(x_i,p_i)$ is a minimum, we obtain 
	\begin{align} 
\frac{\gamma_i }{2 \pi  \hbar L}  \left( \frac{L}{ t} \right)^{\alpha_i+1} \zeta\left( -\alpha_i,  \frac{x-x_i-p_it}{L} -\left\lfloor \frac{x-x_i-p_it}{L} \right\rfloor \right) ,
   \end{align}
and if it is a maximum, we obtain
\begin{align}
	\frac{\gamma_i }{2 \pi  \hbar L}  \left( \frac{L}{ t} \right)^{\alpha_i+1} \zeta\left( -\alpha_i,  \frac{x_i+p_it-x}{L} -\left\lfloor \frac{x_i+p_it-x}{L} \right\rfloor \right),
\end{align}
where $\alpha_i$ and $\gamma_i$ are defined through the expansion around the extremum $(x_i,p_i)$ similar to \eqref{profile-edge}.
}.
Therefore, again the flattest extremum is relevant to determine the exponent of the late time relaxation.

\paragraph{Other local observables:}

We can apply the calculation of $\rho(x,t)$ to other local observables.
We consider the following quantity 
\begin{align}
	F(x,t) \equiv \int_{-\infty}^\infty \frac{dp}{2\pi \hbar} f(p,x) u(x,p,t) ,
	\label{F-local}
\end{align}
where $f(p,x)$ is a smooth function.
Actually various observables in our model are given in this form. 
For example, if we set $f(x,p)=1$ and $f(x,p)=p$, we obtain the particle density $\rho(x,t)$ and the velocity field $v(x,t)$ through (\ref{densities}).
If we take $e^{-ip(x-y)}$, we obtain the two point function $G_F(x,y,t) \equiv \langle \psi^\dagger (y,t)   \psi(x,t) \rangle$ as 
\begin{align}
	G_F(x,y,t)&   = \sum_{n=0}  \frac{(x-y)^n}{n!} \langle \psi^\dagger (y,t)    \partial_y^n \psi(y,t) \rangle \nonumber \\
	&=  \sum_{n=0}  \frac{(x-y)^n}{n!} \int_{-\infty}^\infty \frac{dp}{2\pi} (-i p)^n u(y,p,t) = \int_{-\infty}^\infty \frac{dp}{2\pi } e^{-i(x-y)p} u(y,p,t) .
	\label{2pt-def}
\end{align}

We calculate $F(x,t)$ for the droplet in the case of Figure \ref{fig:quadratic}. 
We substitute \eqref{u-resum} to \eqref{F-local} and perform the $p$ integral similar to \eqref{integral-p}.
There, since the extrema $p_1$ and $p_2$ dominate in the integrals, we can approximate $f(p,x)=f(p_i,x)$. Thus we obtain
\begin{align}
	F(x,t) =& \int_{p_1}^{p_2} \frac{dp}{2\pi \hbar} f(p,x) \frac{x_{0+}(p)-x_{0-}(p)}{L} \nonumber \\
&		+\frac{\gamma_1  }{2 \pi  \hbar L} f(p_1,x)  \left( \frac{L}{ t} \right)^{\alpha_1+1} \zeta\left( -\alpha_1,  \frac{x-x_1-p_1t}{L} -\left\lfloor \frac{x-x_1-p_1t}{L} \right\rfloor \right) \nonumber \\
&	+
	\frac{\gamma_2  }{2 \pi  \hbar L} f(p_2,x) \left( \frac{L}{ t} \right)^{\alpha_2+1} \zeta\left( -\alpha_2,  \frac{x_2+p_2t-x}{L} -\left\lfloor \frac{x_2+p_2t-x}{L} \right\rfloor \right) \nonumber \\
&		+\frac{\gamma^-_1 }{2 \pi  \hbar L} f(p_1,x)  \left( \frac{L}{ t} \right)^{\alpha^-_1+1} \zeta\left( -\alpha^-_1,  \frac{x-x_1-p_1t}{L} -\left\lfloor \frac{x-x_1-p_1t}{L} \right\rfloor \right) \nonumber \\
&	+
\frac{\gamma^-_2 }{2 \pi  \hbar L} f(p_2,x) \left( \frac{L}{ t} \right)^{\alpha^-_2+1} \zeta\left( -\alpha^-_2,  \frac{x_2+p_2t-x}{L} -\left\lfloor \frac{x_2+p_2t-x}{L} \right\rfloor \right)+\cdots,
\label{F-exact}
\end{align}
at large $t$.
Again it shows the power law relaxation.
Note that the exponent is independent of $f(p,x)$.
Thus the local observables described by \eqref{F-local} will show the same power law damping. This is an important finding as it demonstrates some sense of universality, i.e., various relevant quantities, i.e., density field, velocity field, and two-point correlators, have the same long time behaviour. 

\subsection{Particles released from a potential}
\label{sec:release}

To see the relaxation process more concretely, we consider the time evolution of the particles released from a potential $V=cx^{2m}$.
Suppose $N$ particles are trapped by this potential and stay at the ground state. 
Then the initial profile  $x_{0\pm}(p)$ is given by
\begin{align}
\epsilon_F = \frac{p^2}{2}+cx^{2m} \Rightarrow
x_{0\pm}(p)= \pm \left(\frac{p_0^2-p^2}{2c} \right)^{\frac{1}{2m}}, \qquad p_0^2 \equiv 2 \epsilon_F .
\label{trapped-pot}
\end{align}
Here $\epsilon_F $ is the Fermi energy which is determined by the constraint (\ref{ini-constr}) as
follows
\begin{align}
1= &  2 \int_{-p_0}^{p_0} \frac{dp}{2\pi}  \left(\frac{p_0^2-p^2}{2c} \right)^{\frac{1}{2m}} 
=\frac{ p_0}{ \sqrt{\pi}} \left( \frac{p_0^2}{2c} \right)^{\frac{1}{2m}}  \frac{\Gamma\left( 1+\frac{1}{2m}\right) }{\Gamma\left( 1+\frac{1}{2m}+\frac{1}{2}\right)} \nonumber \\
&\Rightarrow p_0=\left( \sqrt{\pi} (2c)^{\frac{1}{2m}} \frac{\Gamma\left( 1+\frac{1}{2m}+\frac{1}{2}\right)}{\Gamma\left( 1+\frac{1}{2m}\right) } \right)^{\frac{m}{m+1}}.
\end{align}
At $t=0$, we suddenly turn off this potential ($V \to 0$). 
Then the particles start evolving according to \eqref{periodic-u}.
At large $t$, the particle density becomes
\begin{align}
\rho(x,t)=&\frac{N}{L} + \frac{ 1}{\pi \hbar }    
  \left(\frac{L p_0}{ c }\right)^{\frac{1}{2m}}   \left(\frac{1}{t}\right)^{1+\frac{1}{2m}} 
  \Biggl[ \zeta\left( - \frac{1}{2m}, \frac{p_0t-x}{L}-\left\lfloor \frac{p_0t-x}{L} \right\rfloor \right)
   \nonumber \\ &\qquad \qquad \qquad \qquad \qquad\qquad  
   +\zeta\left( - \frac{1}{2m}, \frac{p_0t+x}{L}-\left\lfloor \frac{p_0t+x}{L} \right\rfloor \right) 
  \Biggr]
  + \cdots    ,
  \label{damp-trap}
\end{align}
through \eqref{F-exact}.
Therefore, the particle density $\rho(x,t)$ shows a power law relaxation $\sim t^{-\frac{2m+1}{2m}}$.
Note that this result is consistent with the previous study \cite{1742-5468-2013-09-P09025} which investigated the $m=1$ case and observed the damping of the density as $\sim t^{-3/2}$.

\subsection{Particles released from a box}
\label{sec:box}

As the second example, we consider the evolution of the particles released from a box (infinite potential walls). 
Suppose $N$ particles are confined in a region $(-x_0 \le x \le x_0)$ by infinite potential walls and stay at the ground state.
Then the profile of $x_{0\pm}(p)$ is given by
\begin{align}
x_{0\pm}(p)= \pm x_0 \theta(p+p_0)\theta(p_0-p), \qquad p_0=\frac{\pi}{2x_0},
\label{trapped-box}
\end{align}
where $p_0$ is determined so that it satisfies the constraint (\ref{ini-constr}).
At $t=0$, we remove the potential and release the particles on the circle.

In this case, we cannot apply the assumption \eqref{profile-edge}. 
However we can compute the density exactly.
By substituting $x_{0\pm}(p)$ (\ref{trapped-box}) to \eqref{rho-resum}, we obtain
\begin{align}
	\rho(x,t)    =&
	 \frac{N}{L} +   \sum_{k\neq 0}^{\infty} \int_{-p_0}^{p_0} \frac{ dp}{2\pi^2 \hbar k}        
	\left(e^{ \frac{2\pi i k}{L}\left(x_0+pt-x\right) } -
	e^{ \frac{2\pi i k}{L}\left(-x_0+pt-x\right)  }\right) \nonumber \\
	= &
	 \frac{N}{L}  + \frac{L }{2\pi \hbar t} \left\{ \zeta\left(-1, \frac{p_0t+x+x_0}{L} - \left\lfloor \frac{p_0t+x+x_0}{L} \right\rfloor \right)
	-\zeta\left(-1, \frac{p_0t+x-x_0}{L}- \left\lfloor \frac{p_0t+x-x_0}{L} \right\rfloor \right) \right\} \nonumber \\
	&+ \frac{L }{2\pi \hbar t} \left\{ \zeta\left(-1, \frac{p_0t-x+x_0}{L} - \left\lfloor \frac{p_0t-x+x_0}{L} \right\rfloor \right)
	-\zeta\left(-1, \frac{p_0t-x-x_0}{L}- \left\lfloor \frac{p_0t-x-x_0}{L} \right\rfloor \right) \right\} .
\end{align}
Here we use the relation between the $\zeta$ function and the Bernoulli polynomial $\zeta(-n,x)= -\frac{B_{n+1}(x)}{n+1}$, $(n=1,2,\cdots)$ and $B_2(x) = x^2-x-\frac{1}{6}$, and we obtain
\begin{align}
\rho(x,t)=&\frac{N}{L}+ \frac{L }{4\pi \hbar t}
  \left\{  \frac{p_0t+x+x_0}{L} - \left\lfloor \frac{p_0t+x+x_0}{L} \right\rfloor -\left( \frac{p_0t+x+x_0}{L} - \left\lfloor \frac{p_0t+x+x_0}{L} \right\rfloor \right)^2
 \right\}
  \nonumber \\
& - \frac{L }{4\pi \hbar t}
\left\{  \frac{p_0t+x-x_0}{L} - \left\lfloor \frac{p_0t+x-x_0}{L} \right\rfloor -\left( \frac{p_0t+x-x_0}{L} - \left\lfloor \frac{p_0t+x-x_0}{L} \right\rfloor \right)^2
\right\}
\nonumber \\
&+ \frac{L }{4\pi \hbar t}
\left\{  \frac{p_0t-x+x_0}{L} - \left\lfloor \frac{p_0t-x+x_0}{L} \right\rfloor -\left( \frac{p_0t-x+x_0}{L} - \left\lfloor \frac{p_0t-x+x_0}{L} \right\rfloor \right)^2
\right\}
\nonumber \\
& - \frac{L }{4\pi \hbar t}
\left\{  \frac{p_0t-x-x_0}{L} - \left\lfloor \frac{p_0t-x-x_0}{L} \right\rfloor -\left( \frac{p_0t-x-x_0}{L} - \left\lfloor \frac{p_0t-x-x_0}{L} \right\rfloor \right)^2
\right\}.
 \label{rho-box}
\end{align}
We plot this result in Figure \ref{fig-box}.
Thus $\rho(x,t)$ shows a power law relaxation $\sim 1/t$.
Here this relation may be interpreted as the $\alpha \to 0$ case in \eqref{density-decay}.

\begin{figure}[H]
\centering
\begin{minipage}{.5\textwidth}
 \centering
        \includegraphics[scale=1]{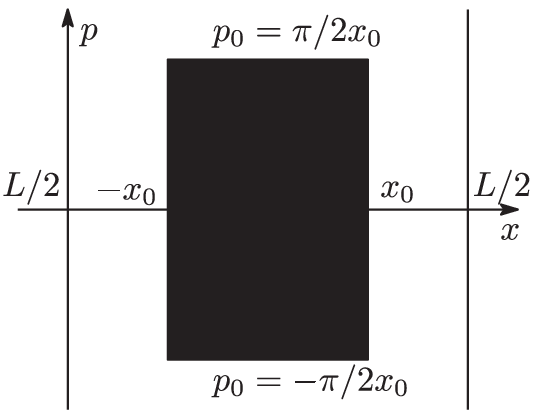}
    \end{minipage}%
    \begin{minipage}{0.5\textwidth}
        \centering
        \includegraphics[scale=0.5]{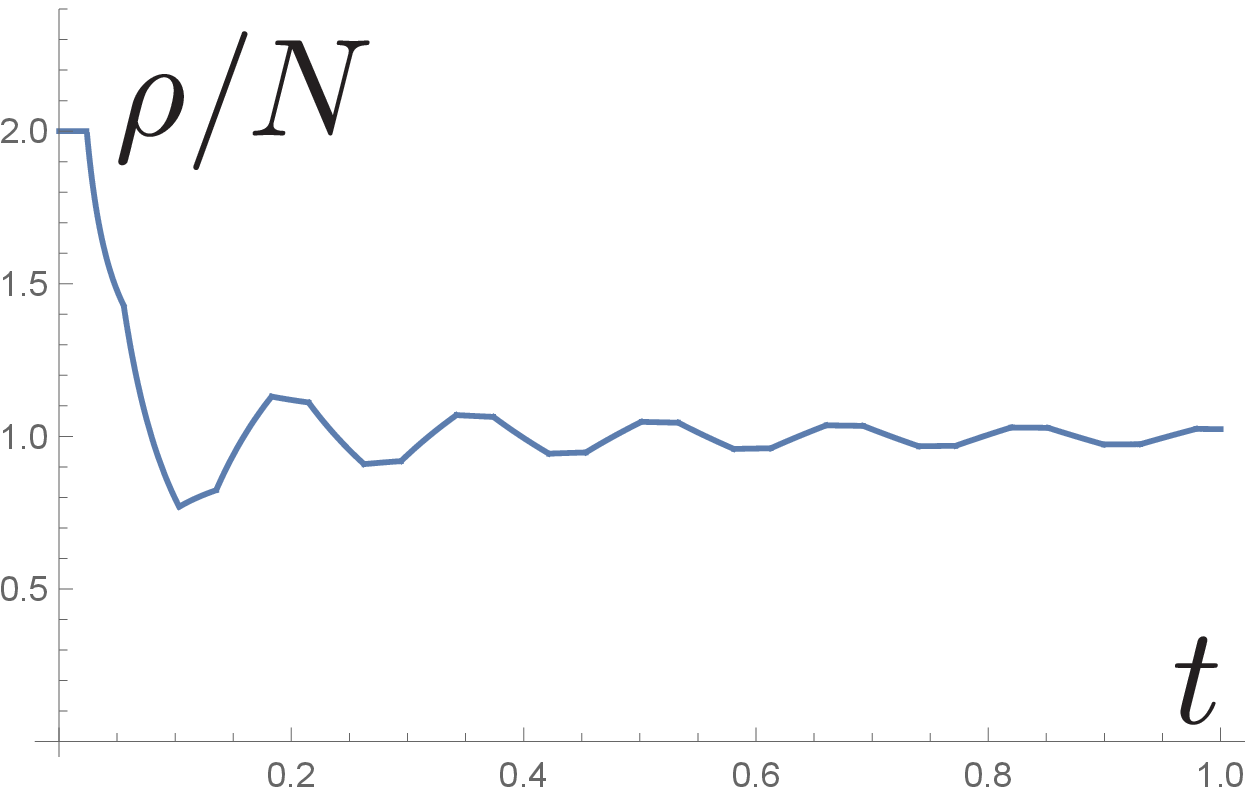}
    \end{minipage}
\caption{ (Left) Droplet which is confined by the infinite potential walls at $x=\pm x_0$ (box potential). 
	(Right) Time evolution of the density $\rho(x,t)$ \eqref{rho-box} at $x=0.4 x_0$. 
	We take $L=2x_0=1$.
	We remove the infinite potential walls at $t=0$ suddenly and release the Fermions.
We can explicitly see that the density relaxes to a steady state $\rho/N=1/L=1$.}
\label{fig-box}
\end{figure}

\section{Particle motions in the external potential (the $V \neq 0$ case)}\label{v-not-zero}

We investigate the time evolution in the $V \neq 0$ case.
Again we put the Fermions on a periodic circle but the arguments in this section can be applied to the relaxation of the Fermions confined in a potential without introducing a circle.

In the phase space, each particle moves along the constant $E$ slices, where $E$ is the energy of the individual particles.
By regarding this point, we can solve the phase space hydrodynamic equations (\ref{u-eom}) and (\ref{u-constr}) as
 \begin{align}
u(x,p,t)=& \theta \left( t - \int_{x_{0+}(E(x,p))}^x \frac{dy}{p(E(x,p),y)} \right) \theta \left(
 \int_{x_{0-}(E(x,p))}^x \frac{dy}{p(E(x,p),y)} -t
\right)   ,\nonumber \\
& E(x,p) \equiv \frac{p^2}{2} +V(x)  , \qquad p(E,y)\equiv \sqrt{2(E-V(y))}.
\label{u-x-E}
\end{align}
Here we have not taken into account the periodicity of the motion and will do it later.
In this equation, $x_{0\pm}(E(x,p))$ denotes the locations of the boundary of the droplet at $t=0$ at a given energy $E(x,p)$.
Therefore, 
\begin{align*}
\int_{x_{0\pm }(E(x,p))}^x \frac{dy}{p(E(x,p),y)},
\end{align*}
represents a ``traveling time" which  a particle at $(x,p)$ with energy $E(x,p)$ spends for traveling from $x=x_{0\pm }(E(x,p))$  to $x$ (related to the time-of-flight coordinate).
Thus Eq. (\ref{u-x-E}) simply says that the droplet is 1 only if 
\begin{align*}
\int_{x_{0+ }(E(x,p))}^x \frac{dy}{p(E(x,p),y)} \le t \le \int_{x_{0- }(E(x,p))}^x \frac{dy}{p(E(x,p),y)}.
\end{align*}
 This result is physically reasonable and we can also check that Eq. (\ref{u-x-E}) satisfies (\ref{u-eom}) and (\ref{u-constr}).

\begin{figure}
  \begin{center}
    \includegraphics[scale=0.7]{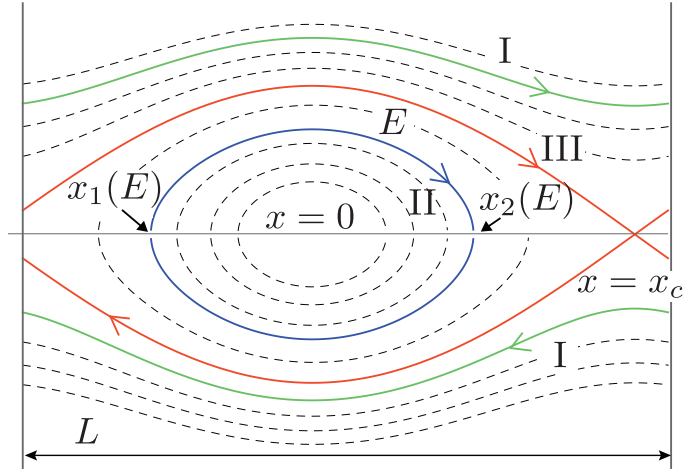}
\caption{Three types of particle motions in the phase space.
	The dotted lines are the constant energy slices.
(I) Right (left) moving motion (the green curves).
(II) Confined motion due to the potential (the blue curve).
(III) Separatrix of these two motions (the red curves).
}
   \label{fig-motions}
  \end{center}
\end{figure}

Note that there are three types of particle motions in this system: (I) right (left) moving forever, (II) trapped by the potential, and (III) the separatrix of these motions.
See Figure \ref{fig-motions}.
(In the $V=0$ case, only the case I appears.) 
We should treat $u(x,p,t)$ separately by regarding these three motions.

For simplicity, we assume that the potential $V(x)$ has only one minimum and one maximum, and we set $x=0$ for the minimum and $x=x_c$ for the maximum respectively.
Also we set $V(0)=0$ and $V(x_c)=E_c$.
Then, if $E>E_c$, the particle motion is type I, and,  if $0<E<E_c$, the particle motion is type II, and $E=E_c$ corresponds to type III.   

Now we consider the periodicity of the motions.
In the case I, because of the periodicity of the circle, the particle with energy $E$ moves with a periodicity
\begin{align}
T_{\text{I}}(E) = \int_0^L \frac{dy}{p(E,y)}.
\end{align}
In the case II, the particle shows a periodic motion with a period
\begin{align}
T_{\text{II}}(E) = 2 \int_{x_1(E)}^{x_2(E)} \frac{dy}{p(E,y)},
\end{align}
where $x_i(E)$ $(i=1,2)$ denotes the two turning points at which $p(E,x_i)=0$.
See Figure \ref{fig-motions}.
On the other hand, in the case III, the particles just approach $x=x_c$ by spending an infinite amount of time, and they do not show periodicity.
Related to this, as $E$ approaches $E_c$, the period $T(E)$ diverges logarithmically.
Indeed, by assuming $V(x) \simeq E_c - \frac{\omega_0^2}{2}(x-x_c)^2 $, we can estimate the periodicity for $E >E_c$ as
\begin{align}
T_{\text{I}}(E) = &\int_0^L  \frac{dy}{p(E,y)} \sim  \int_{x_c-\alpha}^{x_c+\alpha}  \frac{dy}{\sqrt{2(E-E_c)+\omega_0^2 (y-x_c)^2}} \nonumber \\
= & 2 \ \int_{0}^{\alpha} \frac{dz}{\sqrt{2(E-E_c)}}  \frac{1}{\sqrt{1+\frac{\omega_0^2 z^2}{2(E-E_c)}}},
\end{align}
where $\alpha$ is a some scale and we have used $z=y-x_c$.
Then by defining $ z= \frac{\sqrt{2(E-E_c)}}{\omega_0} \sinh \theta $, through the $\theta$ integral, we obtain
\begin{align}
T_{\text{I}}(E)\sim \frac{1}{\omega_0} \log \left( \frac{\alpha \omega_0}{\sqrt{E-E_c}}\right),
\label{T-Ec}
\end{align}
for $E \sim E_c$ and it diverges logarithmically. We obtain a similar result for $T_{\text{II}}(E)$ in the case II ($E<E_c$).
This logarithmic divergence will be important when we discuss the power law relaxation \footnote{The particle motion near the critical point $x=x_c$ saturates the bound on chaos \cite{Maldacena:2015waa}, once we turn on the quantum effect \cite{Morita:2018sen}. }.

By taking into account these periodicities, the Wigner distribution (\ref{u-x-E}) for the case I becomes
\begin{align}
u(x,p,t)=&  \sum_m  \theta \left( t-mT_{\text{I}}(E(x,p)) - \int_{x_{0+}(E(x,p))}^x \frac{dy}{p(E(x,p),y)}  \right)  \nonumber \\
& \qquad  \qquad \qquad 
\times \theta \left(
 \int_{x_{0-}(E(x,p))}^x \frac{dy}{p(E(x,p),y)} -t + mT_{\text{I}}(E(x,p))
\right)   \nonumber \\
=& \frac{1}{T_{\text{I}}(E(x,p))} \int_{x_{0-}(E(x,p))}^{x_{0+}(E(x,p))} \frac{dy}{p(E(x,p),y)}  \nonumber \\
&+ \sum_{k=1}^\infty \frac{2}{\pi k} \sin \left( \frac{2\pi k}{T_{\text{I}}(E(x,p))} 
\frac{1}{2} \int_{x_{0-}(E(x,p))}^{x_{0+}(E(x,p))} \frac{dy}{p(E(x,p),y)} \right) \nonumber \\
&\times \cos \left(
\frac{2\pi k}{T_{\text{I}}(E(x,p))} \left(t- \frac{1}{2}\left( \int_{x_{0+}(E(x,p))}^x \frac{dy}{p(E(x,p),y)}+\int_{x_{0-}(E(x,p))}^x \frac{dy}{p(E(x,p),y)}\right) \right) 
\right),
\label{u-result-V}
\end{align}
where we have used the Poisson summation formula (\ref{Poisson}).
Here the first term is time independent and it would describe a late time steady state, while the second time dependent terms will describe the relaxation\footnote{Gaussian potential $V=\frac{\omega^2}{2}x^2$ is an exception \cite{PhysRevA.97.033609}. In this case, the droplet rotates in the phase space with a constant periodicity $2\pi/ \omega$ forever and does not show a relaxation. }.
We will have the same formula for the case II also by replacing $T_{\text{I}} \to T_{\text{II}}$.

By using this result, we can evaluate various observables in this system.
As an example,  we investigate the particle density $\rho(x,t)$.
From (\ref{densities}), we obtain
\begin{align}
\rho(x,t)
 =& \int_{-\infty}^{\infty}  \frac{dp}{2\pi \hbar} u(x,p,t)
  =  \frac{1}{2\pi \hbar} \int_{0}^{\infty} \frac{dE}{ p(E,x)} u(x,p(E,x),t)
 +\frac{1}{2\pi \hbar} \int_{0}^{\infty} \frac{dE}{ p(E,x)} u(x,-p(E,x),t)
  \nonumber \\
 = &
    \frac{1}{2\pi \hbar} \int_{E_c}^{\infty} \frac{dE}{ p(E,x)} u(x,p(E,x),t)
    +\frac{1}{2\pi \hbar} \int_{0}^{E_c} \frac{dE}{ p(E,x)} u(x,p(E,x),t) \nonumber \\
& +\frac{1}{2\pi \hbar} \int_{0}^{E_c} \frac{dE}{ p(E,x)} u(x,-p(E,x),t)
+\frac{1}{2\pi \hbar} \int_{E_c}^{\infty} \frac{dE}{ p(E,x)} u(x,-p(E,x),t).
\label{rho-V-1}
 \end{align}
Here we have used $dp = dE/p(E,x) =dE/\sqrt{2(E-V(x))}$. (We have taken $p(E,x)=\sqrt{2(E-V(x))}$ positive.)
In this equation, the first term is for the type I particle motion (right mover), the second and third terms are for type II and the fourth one is for type I (left mover).
The first term can be calculated by using (\ref{u-result-V})
\begin{align}
&  \frac{1}{2\pi \hbar} \int_{E_c}^{\infty} \frac{dE}{ p(E,x)} u(x,p(E,x),t)  \nonumber \\
  =
&\frac{1}{2\pi \hbar} \int_{E_c}^{\infty}  \frac{dE}{ p(E,x)}  \frac{1}{T_{\text{I}}(E)} \int_{x_{0-}(E)}^{x_{0+}(E)} \frac{dy}{p(E,y)} \nonumber \\
&+ \sum_{k=1}^\infty \int_{E_c}^{\infty}  \frac{dE}{ p(E,x)}   \frac{1}{\pi^2 \hbar k} \sin \left( \frac{2\pi k}{T_{\text{I}}(E)} 
\frac{1}{2} \int_{x_{0-}(E)}^{x_{0+}(E)} \frac{dy}{p(E,y)} \right) \nonumber \\
& \qquad \times \cos \left(
\frac{2\pi k}{T_{\text{I}}(E)} \left(t- \frac{1}{2}\left( \int_{x_{0+}(E)}^x \frac{dy}{p(E,y)}+\int_{x_{0-}(E)}^x \frac{dy}{p(E,y)}\right) \right) 
\right).
\label{rho-V}
\end{align}
We will have similar expressions for the other three terms in (\ref{rho-V-1})

For large $t$, since the time dependent terms would highly oscillate and become sub-dominant, we would obtain the late time behaviour,
\begin{align}
\rho(x,t)
=&\frac{1}{2\pi \hbar} \int_{0}^{\infty}  \frac{dE}{ p(E,x)}  \frac{1}{T_{\text{I}}(E)} \int_{x_{0-}(E)}^{x_{0+}(E)} \frac{dy}{p(E,y)} +  \text{(contributions from left mover)}, \qquad (t \to \infty).
\label{rho-steady-V}
\end{align}
Here the contributions from the left mover is given by the same expression to the first term but $x_{0\pm}(E)$ is taken as the one with $p<0$.
In contrast to the $V=0$ case, this density depends on $x$. 
This is natural since the potential depends on $x$.

\subsection{Power law relaxation}\label{power-law-v-nonzero}

\begin{figure}
	\begin{center}
		\includegraphics[scale=0.7]{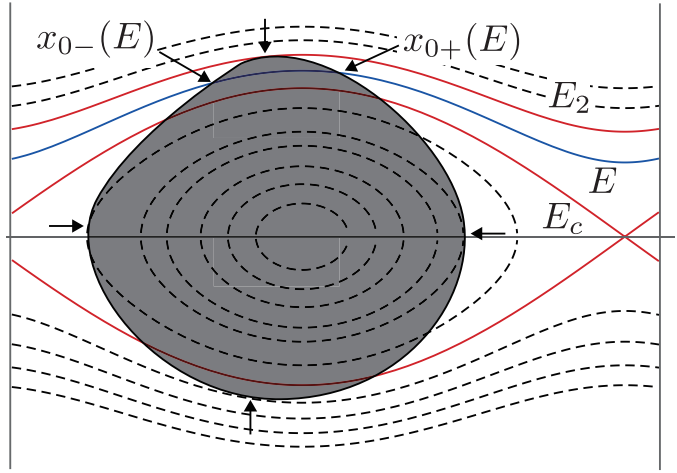}
\caption{Initial profile of a droplet in the phase space.
	$E_2$ is the maximal energy of the particle in the droplet.
	$E_c$ is the energy of the particle on the separatrix.
	We define $x_{0\pm}(E)$ as the boundaries of the droplet which intersect with the constant energy curve at energy $E$. 
	The points at which $\partial_E x_{0\pm}(E)$ diverge determine the power of the late time damping.
	The small arrows indicate such points.
}
\label{fig-evolution}
	\end{center}
\end{figure}

To see how the system relaxes to the steady state (\ref{rho-steady-V}), we evaluate the time dependent terms of $\rho(x,t)$ in (\ref{rho-V}).
By regarding the result in the $V=0$ case, we presume that the relevant contributions to the large $t$ behaviour will arise from the location of the initial profile which hits the local maximum or minimum energy slices.
See Figure \ref{fig-evolution}.
There, $x_{0+}(E)$ and $x_{0-}(E)$ meet and show a singularity as a function of $E$.
($\partial_E x_{0 \pm}(E)$ may diverge.)
Besides, the droplet around $E=E_c$ might show some singular behaviour and we need to evaluate whether it gives a relevant contribution or not.

First we evaluate the $E$ integral \eqref{rho-V} in $\rho(x,t)$ near the maximum energy, which we define $E_2$.
We assume that $E_2>E_c$ and $x_{0\pm}(E) $ near $E =E_2$ behave as
 \begin{align} 
 x_{0+}(E) = & x_2+ \gamma_+ (E_2-E)^{\alpha_+} + \cdots, \qquad
 x_{0-}(E) = x_2- \gamma_- (E_2-E)^{\alpha_-}  + \cdots ,
 \end{align}
where $x_2:=x_{0\pm}(E_2)$, and $\gamma_{\pm }$ and $\alpha_\pm$ are positive constants.
Then we can approximate 
\begin{align}
 \int_{x_{0-}(E)}^{x_{0+}(E)} \frac{dy}{p(E,y)}  \simeq & \frac{1}{p(E_2,x_2)} \left(
\gamma_+ (E_2-E)^{\alpha_+}+\gamma_- (E_2-E)^{\alpha_-}
 \right) 
  \simeq  \frac{1}{p(E_2,x_2)} 
\gamma (E_2-E)^{\alpha},
 \label{x+-E2}
\end{align}
where $\alpha = \min (\alpha_+, \alpha_-)$ and $\gamma$ is $\gamma_+$ or $\gamma_-$ correspondingly.
(If the droplet is smooth, $\alpha_+= \alpha_-$ and $\gamma_+=\gamma_-=\gamma/2$.)
Besides we can expand
\begin{align}
T_{\text{I}}(E) =T_{\text{I}}(E_2) -(E_2-E) \partial_E T_{\text{I}}(E_2)+ \cdots, \qquad
\partial_E T_{\text{I}}(E_2) =-  \int_0^L \frac{dy}{p(E_2,y)^3} ,
\end{align}
where $\partial_E T_{\text{I}}(E_2) $ is negative. 
As we did in the $V=0$ case, we change the integral contour to the imaginary direction in (\ref{rho-V}).
To do so, we take $E=E_2+i \eta$ and do the integral with respect to $\eta$ $(0 \le \eta \le \infty)$. Then,  for large $t$, we obtain the contribution from $E \sim E_2$ as 
\begin{align}
& \int^{E_2}  \frac{dE}{ p(E,x)}   \sin \left( \frac{2\pi k}{T_{\text{I}}(E)} 
\frac{1}{2} \int_{x_{0-}(E)}^{x_{0+}(E)} \frac{dy}{p(E,y)} \right) \nonumber \\
& \qquad \times \cos \left(
\frac{2\pi k}{T_{\text{I}}(E)} \left(t- \frac{1}{2}\left( \int_{x_{0+}(E)}^x \frac{dy}{p(E,y)}+\int_{x_{0-}(E)}^x \frac{dy}{p(E,y)}\right) \right) 
\right) \nonumber \\
= & - \int_{0}^{\infty} \frac{ i d\eta}{ p(E_2,x)}  \frac{2\pi k}{T_{\text{I}}(E_2)} 
 \frac{ \gamma \left(-i \eta\right)^{\alpha}}{2p(E_2,x_2)} \nonumber \\
&\quad \times
\exp \left(
-\frac{2\pi k \left(-\partial_E T_{\text{I}}(E_2)\right)}{T_{\text{I}}(E_2)^2} t \eta
+\frac{2\pi k i}{T_{\text{I}}(E_2)}\left(t-  \frac{1}{2}\left( \int_{x_{0+}(E_2)}^x \frac{dy}{p(E_2,y)}+\int_{x_{0-}(E_2)}^x \frac{dy}{p(E_2,y)}\right)  
\right)\right)+\text{c.c.}+ \cdots \nonumber \\
\sim & \frac{1}{t^{1+ \alpha} }.
\label{relax-V}
\end{align}
Similar contributions will arise from other singular points on the initial droplet too.
Therefore we obtain a power law relaxation even in the $V\neq 0$ case.

To confirm this power law relaxation, we need to evaluate the contribution around $E =E_c$ and see whether it breaks our result or not. We consider it in the case I.
As $E$ approaches $E_c$, $T_{\text{I}}(E)$ shows a logarithmic divergence as in (\ref{T-Ec}), and the $E$ integral in (\ref{rho-V}) near $E_c$ can be estimated as
\begin{align}
&  \int_{E_c} \frac{dE}{ p(E,x)}\sin \left( \frac{2\pi k}{T_{\text{I}}(E)} 
\frac{1}{2} \int_{x_{0-}(E)}^{x_{0+}(E)} \frac{dy}{p(E,y)} \right) \nonumber \\
& \qquad \times \cos \left(
\frac{2\pi k}{T_{\text{I}}(E)} \left(t- \frac{1}{2}\left( \int_{x_{0+}(E)}^x \frac{dy}{p(E,y)}+\int_{x_{0-}(E)}^x \frac{dy}{p(E,y)}\right) \right) 
\right) \nonumber \\
\sim &  \int_{E_c} \frac{dE}{ p(E_c,x)}   \frac{1}{-\log \left(E-E_c \right)}\int_{x_{0-}(E_c)}^{x_{0+}(E_c)} \frac{dy}{p(E_c,y)} \exp \left(
\frac{it}{-\log \left(E-E_c \right)} + \cdots
\right) + \text{c.c.} \nonumber \\
\sim &  \int_{0}^{\infty} id\eta   \frac{1}{-\log \left( i \eta \right)} \exp \left(
\frac{it}{-\log i \eta} + \cdots
\right) + \text{c.c.} \nonumber, 
\end{align}
where have used $E=E_c+ i \eta$ . We define a new variable $w$ via  $w  = - \frac{t^{1/2}}{\log \eta}$, and then the integral becomes
\begin{align}
\sim &  \int_{0}^{\infty} d w    \frac{1}{w} \exp \left( -t^{1/2} \left( \frac{1}{w}-iw \right) \right) + \text{c.c.} \nonumber.
\end{align}
Through the saddle point approximation, we see that this integral is exponentially suppressed at large $t$.
Thus the contributions around $E=E_c$ will be irrelevant. Therefore we conclude that the dominant contribution for the relaxation at large $t$ arises from the singular $x_{0\pm}(E)$ at $t=0$ as shown in (\ref{relax-V}) and the relaxation always obeys the power law.

\subsection{Example:  Sudden trap from $V=0$ to $ V=V_0 \cos \left( \frac{2 \pi x}{L} \right)$}
\label{sec-cos}

\begin{figure}
	\begin{center}
		\includegraphics[scale=0.5]{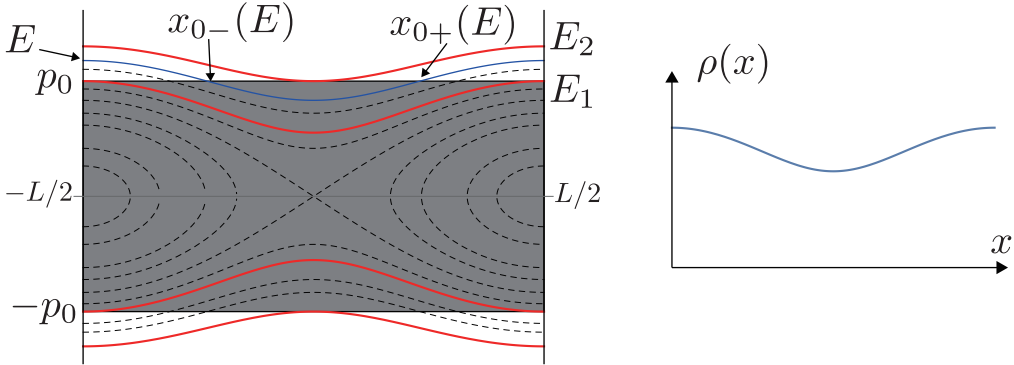}
		\caption{(Left) Initial profile of the droplet in the phase space in the case of the quench from $V=0$ to $ V=V_0 \cos \left( \frac{2 \pi x}{L} \right)$.
			The shadow is the droplet at $t=0$. 
			The red curves are the constant energy slices for $E=E_1$ and $E_2$ which touch the surface of the droplet.
			(Right) Density of the Fermions at $t= \infty$.
		}
		\label{fig-cos}
	\end{center}
\end{figure}

As an example, we consider the following quench procedure.
We start from a system on a circle with a periodicity $L$ and set $V=0$.
We consider the ground state of the $N$ Fermions on this circle and, at $t=0$, turn on a potential $V=V_0 \cos\left( \frac{2 \pi x}{L} \right)$, and see the evolution\footnote{Quantum quench from $V=V_1 \cos\left( \frac{2 \pi x}{L} \right)$ to $V_2 \cos\left( \frac{2 \pi x}{L} \right)$ has been studied by two of the authors \cite{Mandal:2013id}. It includes exact computations at finite $N$, dynamical phase transitions, and properties near the critical point.}. 

Before the quench, the Fermions are filled up to $p_0 := \pi \hbar N/L$ on the phase space \eqref{u-constr}. 
For simplicity, we tune $0<V_0< p_0^2/2 $ so that the Fermi surface does not cross the separatrix $p=\pm 2\sqrt{V_0}\sin\left(\pi x/L \right)$ after the quench.
See Figure \ref{fig-cos}.
In this case, the motion of the surface of the droplet after the quench is type I with energy between $E_1:=p_0^2/2-V_0$ and $E_2:=p_0^2/2+V_0$.
We also obtain the initial boundaries of the droplet $x_{0\pm}(E)$  as
\begin{align}
	p(E,x)=\pm \sqrt{2\left(E-V_0 \cos\left( \frac{2 \pi x}{L} \right) \right )} \quad \Longrightarrow \quad
	x_{0\pm}(E)= \pm \frac{L}{2\pi} \arccos\left( \frac{E-p_0^2/2}{V_0}\right).
\end{align}
Note that these boundaries have two extrema at $E=E_1$ ($x= L/2$) and $E=E_2$ ($x=0$), and we can expand them as
\begin{align}
	x_{0\pm}(E)= \pm \frac{L}{2\pi} \sqrt{\frac{2(E_2-E)}{V_0}}+ \cdots, \qquad 	x_{0\pm}(E)= \frac{L}{2} \mp \frac{L}{2\pi} \sqrt{\frac{2(E-E_1)}{V_0}}+ \cdots.
\end{align}
By comparing these expansions with \eqref{x+-E2}, we can read off $\alpha=1/2$.
Therefore we conclude that the time dependent terms damp as $t^{-3/2}$ at large $t$ from \eqref{relax-V}.

Note that, from \eqref{rho-steady-V}, we can compute the particle density at the steady state.
The result is plotted in Figure \ref{fig-cos}.
There, some of the particles are trapped around $x= \pm L/2$.

In the case of $V_0 > p_0^2/2$, we will see the same power law damping $t^{-3/2}$, although the calculations are slightly involved.

\section{GGE conjecture and Entropy production}\label{sec:GGE}

As we have shown in the previous sections, the system evolves to the steady state through the power law relaxation.
There, the Wigner distribution (\ref{u-result-V}) with  $E>E_c$ and $p>0$ effectively reduces to
\begin{align}
u(x,p,t) \to \frac{1}{T_{\text{I}}(E(x,p))} \int_{x_{0-}(E(x,p))}^{x_{0+}(E(x,p))} \frac{dy}{p(E(x,p),y)},
\qquad (t \to \infty).
\label{u-steady}
\end{align}
as far as the local observables \eqref{F-local} are concerned.
 We will show that this reduced distribution function indeed agrees  with the GGE conjecture.
(Generalizations to $0<E<E_c$ or $p<0$ cases are straightforward.)

GGE predicts that, if an integrable system evolves to a steady state through the time evolution, the steady state is approximately described by a density matrix \cite{PhysRevLett.98.050405}
\begin{align}
	\varrho_{\rm GGE}= \frac1{Z_{\rm GGE}}\exp\left[-\sum_m \mu_m \hat{N}_m \right], \qquad
	Z_{\rm GGE}=   \Tr \exp\left[-\sum_m \mu_m \hat{N}_m \right],
	\label{GGE}
\end{align}
where $\hat{N}_m$ is a conserved charge and $\mu_m$ is the corresponding chemical potential.
The chemical potential is determined by the initial data such that the conserved quantity is correctly obtained by $\langle \hat{N}_m \rangle_{\text{GGE}} = N_m |_{t=0}$ as usual, where $\langle \hat{O} \rangle_{\text{GGE}}$ denotes that we evaluate the expectation value by using the density matrix (\ref{GGE}).
In integrable systems, the number of such conserved charges is typically infinity, and in this sense the density matrix differs from the standard thermal density matrix where the number of the conserved charges is finite. 

Let us test this conjecture in our free Fermion model. 
Since the individual energies of each particles are conserved in our system, the number of the particles at each energy level is a conserved quantity.
We define the number of the particle at the $m$-th level as $N_m$.
(For $E\ge E_c$, we have two independent modes (left and right) for a given energy, and $N_m$ for these modes should be distinguished. Here we consider only the right mode.)
Indeed, in the  semi-classical approximation, we can read off $N_m$ from the initial profile of the droplet through
\begin{align}
N_m  =\frac{1}{T_{\text{I}}(E_m)} \int_{x_{0-}(E_m)}^{x_{0+}(E_m)} \frac{dy}{p(E_m,y)} ,
\end{align}
where $E_m$ is the energy of the $m$-th level. 
Now we consider the GGE density matrix (\ref{GGE}) with given $N_m$ and evaluate  the expectation value of  the Wigner distribution function (\ref{wigner-app}) 
\begin{align}
\langle \hat{u}(x,p) \rangle_{\rm GGE}=&  \int d\eta\ \langle  \hat{\psi}^\dagger(x+\eta/2,t)
\hat{\psi}(x-\eta/2,t)\rangle_{\rm GGE} \exp[i\eta p/\hbar] \nonumber \\
=&\sum_m \int d\eta\ N_m \ \varphi_m^\dagger(x+\eta/2,t)
\varphi_m(x-\eta/2,t) \exp[i\eta p/\hbar]  .
\end{align}
Here we have used the expansion of the second quantized field $\hat{\psi} (x)= \sum_m \hat{c}_m \varphi_m (x) $, where $\hat{c}_m$ is the creation/annihilation operator and $\varphi_m(x)$ is the $m$-th eigenfunction of the Hamiltonian (\ref{H-single}) which satisfies $\hat{h} \varphi_m(x) = E_m \varphi_m(x) $.
We have also used $\langle \hat{c}^{\dagger}_n \hat{c}_m  \rangle_{\rm GGE}= N_m \delta_{nm}$.
Then by using the WKB approximation $\varphi_m(x)= \exp( i \int^x p(E_m,y)dy/ \hbar )/  \sqrt{T(E_m) p(E_m,x)}$, we approximately obtain
\begin{align}
\langle \hat{u}(x,p) \rangle_{\rm GGE} \simeq &\sum_m \int d\eta\ \frac{N_m}{T_{\text{I}}(E_m) p(E_m,x)}e^{\frac{i \eta}{\hbar} \left(p- p(E_m,x)  \right)} 
=   \sum_m  \frac{2\pi \hbar N_m }{T_{\text{I}}(E_m)} \delta(E(x,p)-E_m) \nonumber \\
=&   \int d\tilde{E} \ N_m  \delta(E(x,p)-\tilde{E})=  \frac{1}{T_{\text{I}}(E(x,p))} \int_{x_{0-}(E(x,p))}^{x_{0+}(E(x,p))} \frac{dy}{p(E(x,p),y)},
\label{u-GGE}
\end{align}
in the semi-classical limit.
Here we have used $\sum_m = \int dE \frac{ dN}{dE}$ and $\frac{ dN}{dE}= \frac{T(E)}{2\pi \hbar}$. 
This distribution function agrees with the reduced distribution (\ref{u-steady}) of the actual time evolution and the GGE conjecture in our system is proved.

\subsection{Entropy production}\label{delta-s}

As we have seen, the system at late time can be approximately described by using the GGE distribution function (\ref{u-GGE}).
However the value of this distribution function is not always zero or 1 but between zero and 1. 
It means that the droplet for the steady state is not always white or black but can be ``grey". See Figure \ref{fig:thermalization}.
Hence the constraint $u^2=u$ (\ref{u-constr}) is not satisfied anymore.
It implies that this distribution function does not describe a pure state but a mixed state.
Indeed the von Neumann entropy of the GGE density matrix (\ref{GGE}) becomes\cite{Mandal:2013id,PhysRevLett.98.050405}
\begin{align}
	S_{\text{entropy}}=&- \Tr \rho_{\text{GGE}} \log  \rho_{\text{GGE}} = 
	- \sum_{m=0}^\infty \left[ 
	\frac{e^{-\mu_m}}{1+e^{-\mu_m}} \log \frac{e^{-\mu_m}}{1+e^{-\mu_m}} 
	+ 
	\frac{1}{1+e^{-\mu_m}} \log  \frac{1}{1+e^{-\mu_m}}  
	\right]  
	\nonumber \\
	=& - \sum_{m=0}^\infty \left[  N_m  \log   N_m   
	+ \left( 1- N_m  \right) \log  \left( 1- N_m  \right) 
	\right] \nonumber \\
	=& -\int_0^\infty  d E ~ \frac{T(E)}{2\pi \hbar} \Biggl[  \frac{1}{T(E)} \int_{x_{0-}(E)}^{x_{0+}(E)} \frac{dy}{p(E,y)}
	\log  \left(\frac{1}{T(E)} \int_{x_{0-}(E)}^{x_{0+}(E)} \frac{dy}{p(E,y)} \right)  \nonumber \\
	& \qquad + \left( 1-\frac{1}{T(E)} \int_{x_{0-}(E)}^{x_{0+}(E)} \frac{dy}{p(E,y)} \right) \log  \left( 1-\frac{1}{T(E)} \int_{x_{0-}(E)}^{x_{0+}(E)} \frac{dy}{p(E,y)} \right)  \Biggr] \ 
	,
\end{align}
where we have used $Z_{\rm GGE}=\prod_{m=0}^{\infty} (1+e^{-\mu_m}) $ and $N_m = - \partial_{\mu_m} \log Z_{\rm GGE}=1/(1+e^{\mu_m}) $.
Also we have omitted to distinguish I or II and $p>0$ or $p<0$ .
This entropy is finite and is typically proportional to $1/\hbar \sim N$.
This is consistent with our intuition for $N$ particle system.

This result explains the entropy production of our system.
As the system evolves, the oscillating terms in the Wigner distribution function (\ref{u-result-V}) become less relevant for the local observables \eqref{F-local} and 
the system would be approximately described by the time independent term, which is equivalent to the GGE state (\ref{u-GGE}).
Hence, if we employ the GGE distribution function (\ref{u-GGE}) instead of the original distribution function (\ref{u-result-V}), the system at late time can be regarded as a mixed state and entropy is non-zero.
If we use the original distribution function (\ref{u-result-V}), entropy is of course zero.
Thus, although the time dependent oscillating terms are less relevant for the local observables, they significantly contribute to entropy.

\section{Conclusions}
\label{sec-conclusions}

In this paper, we demonstrated via phase-space hydrodynamics, non-trivial thermalization properties of Fermi gas. Conventional hydrodynamics cannot be used to describe long time behaviour as it develops pathologies when shocks develop. (See the Appendix \ref{sec-collective}). On the other hand the phase-space method has no such pathologies and captures all the long-time behaviour. We obtained the power law exponents (initial condition dependent) for evolution of Fermionic density. Thus we determined the power law exponents after releasing the Fermi gas from various experimentally accessible potentials (such as quadratic traps and box-like \cite{zh0,Navon167,navon17,navon17a,navon17b,almostbox} potentials). We also studied the motion of Fermions in a potential at post-quench stage. Some examples of power law relaxation at long times have been summarized in Table \ref{table:power}. We showed that the long time limit of the Wigner function is in agreement with the GGE conjecture. We also showed explicit computations of  relevant entropy production in the system by computing the von Neumann entropy.

We should note here that the sense in which the word thermalization is
used in our paper is different from that of dissipative hydrodynamics.
For example, in the Appendix \ref{sec-collective}, we derive the hydrodynamic
equations \eq{conventional} of conservative fluid dynamics without any
dissipative term like a viscosity.
As pointed out in Ref~\cite{2008PhRvL.100j0601B}, dephasing may explain the thermalization in many body integrable systems,  and this is the physics content of the mathematical steps
described in Eqs. \eq{periodic-u} to \eq{rho-steady} in our model. 
Particularly the Wigner function is decomposed into the time independent term and damping terms through Poisson summation formula \eqref{Poisson}, and this plays the key role to observe the thermalization.
 It is pertinent to mention here that GGE has been derived from
a notion of ``generalized hydrodynamics'' in several recent papers,
e.g., Ref. \cite{Doyon:2016sqa}. It would be interesting to compare the
approach in these papers with ours.

The results we obtained can, in principle, be realized in cold atomic gases. Our setup requires certain important ingredients most of which have already been realized. These ingredients are listed as follows. \textit{(i) Periodic boundary conditions :} It is to be noted that in some sense periodic boundary conditions are experimentally feasible. For example, there has been a lot of progress on preparing ring shaped or toroidal traps \cite{ringPRL, ringv1,ringv2,ringv3}. These confining potentials were experimentally realized precisely with the aim of mimicking periodic boundary conditions. 
\textit{(ii) Box-like potentials :} Box potentials (often referred to as homogenous potentials) have become experimentally realistic due to recent breakthroughs \cite{zh0,Navon167,navon17b,navon17a,navon17}. In such experimental setups, the effect of inhomogeneity due to inevitable external potentials was overcome.  \textit{Quadratic and quartic traps:} Needless to mention harmonic traps are ubiquitous in cold atomic experiments. Additionally, quartic traps have also been realized \cite{quart0, quart1, quart2}. 
 It is to be noted that if we take a linear combination of the potential, i.e., $V=c_1 x^2+c_2 x^4$ (where $c_1$,$c_2>0$), then the long time behaviour of the particles released from this potential to a circle is given by $t^{-3/2}$ via \eqref{damp-trap}, since the initial profile around $x=0$ is relevant.
If we could tune $c_1=0$, we may observe  $t^{-5/4}$.
  \textit{Sinusoidal traps :}  Quench protocols using sinusoidal traps can be achieved in experiments by turning on or off optical potentials \cite{rmpbloch}.
\textit{Absorption imaging techniques: } As the reader might have noticed, most of our major findings involve density dynamics. Evolution of density profiles can be observed in cold atomic experiments via absorption imaging techniques \cite{kulPRL} at large times. Needless to mention, the Tonks gas has been realized in experiments \cite{paredes2004tonksgirardeau, Kinoshita1125}. Integrating some of the above mentioned experimental capabilities would make it possible to realize the predictions made in the paper.

\section*{Acknowledgements} 

We would like to thank Avinash Dhar, Rajesh Gopakumar, Taro Kimura, R. Loganayagam,
Suvrat Raju, Joseph Samuel, and Spenta Wadia for useful
discussions. G.M. would like to thank participants of the Bangalore
Area Strings meeting (31 July to 02 August 2017), ICTS-TIFR,
Bangalore, for stimulating discussions where part of this work was
presented \cite{GM:2017talk}. The work of G.M is supported in part by
Infosys Endowment for the study of the Quantum Structure of Space
Time. M. K. gratefully acknowledges the Ramanujan Fellowship No.
SB/S2/RJN-114/2016 from the Science and Engineering Research Board
(SERB), Department of Science and Technology, Government of India.
The work of T.~M. is supported in part by Grant-in-Aid for Scientific
Research (No. 15K17643) from JSPS.

\appendix

\section{Hydrodynamics of one-dimensional Fermi fluid and its breakdown}\label{sec-collective}

We will now consider an example of a droplet, which, in the
standard $(x,p)$ coordinate system of the phase space, is a single,
connected, ``quadratic droplet'' (with no folds) (defined in Figure \ref{fig-quadratic}).

\begin{figure}[H]
\begin{center}
\includegraphics[scale=1]{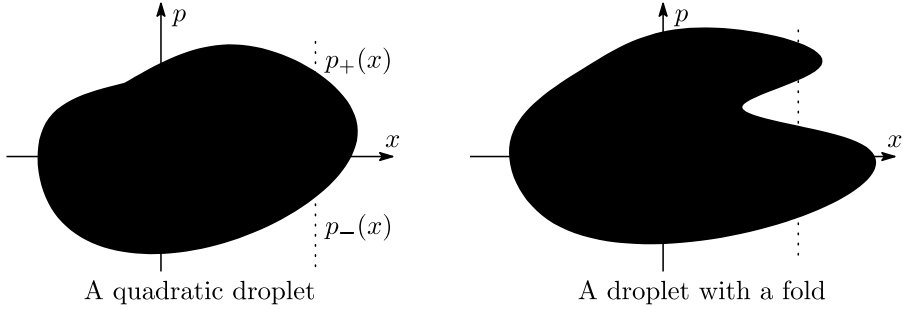}
\caption{Single connected droplets. We call it a ``quadratic droplet''
  if any $x$= constant line intersects the droplet boundary only at
  two points, $p_+(x)$ and $p_-(x)$. 
  We call it a ``fold'' if there is a region in which any $x$= constant line intersects the droplet boundary more than two points. 
  An example of a quadratic droplet is the ground state (the Fermi sea) for $N$ Fermions trapped in a potential $V=x^{2m}$ \eqref{trapped-pot}.
  More complicated examples of
  quadratic droplets can include small fluctuations of such a droplet,
  the Fermi sea for a deformed trap, etc.}
\label{fig-quadratic}
\end{center}
\end{figure}
For ``quadratic droplets'', the phase space density is given by
\begin{align}
u(x,p,t)=  \theta\Big(p_+(x,t)- p\Big)\theta\Big(p- p_-(x,t)\Big).
\label{quadratic-p}
\end{align}
With this, the number density $\rho(x,t)$ and the specific momentum
density $v(x,t)$, defined in \eq{densities}, are given by
\begin{align}
\rho(x,t) & =\f{1}{2\pi \hbar} (p_+(x,t) - p_-(x,t)), \qquad
v(x,t)  = \frac12 (p_+(x,t) + p_-(x,t)).
\label{quad-densities}
\end{align}
The variables $p_\pm(x,t)$ are, therefore, related to
the more physical variables, the densities
\[
p_\pm(x,t)= v(x,t) \pm \pi \hbar \rho(x,t).
\]
It is easy to compute the total energy of a quadratic droplet (we drop
the $t$-dependence since it is a conserved quantity)
\begin{align}
H & = \int \f{dx dp}{2\pi \hbar} h(x,p) u(x,p)
\nonumber\\
&= \int \f{dx}{2\pi \hbar}\left[\f16 \left(p_+(x,t)^3-  p_-(x,t)^3\right) +
V(x) \left(p_+(x,t)-  p_-(x,t)\right)\right]
\nonumber\\
&= \int dx\ \rho(x) \left(\frac12 v(x)^2 + \frac16 \pi^2 \hbar^2 \rho(x)^2 + V(x)\right).
\label{hamiltonian}
\end{align}
The equation of motion \eq{u-eom} now splits into two
independent equations for each variable $p_\pm(x)$ (after
some straightforward manipulation of the $\theta$-functions), \footnote{
The solutions of \eqref{p-plus-minus}  are given as the following parametric set of equations,
\begin{align}
	p_{\pm}(x)=\sqrt{2 (E - V(x))}, \quad
	t=\int_s^x \frac{dy}{\sqrt{2 (E - V(y))}},
	\quad	E= \frac{p_{\pm}^{initial}(s)^2}{2} +V(s)
\end{align}
where $p_{\pm}^{initial}(s)$ is the initial condition. 
See, for example, Ref. \cite{kul1}. }
\begin{align}
\del_t p_+ + \del_x( p_+^2/2 + V(x))=0, \qquad
\del_t p_- + \del_x( p_-^2/2 + V(x))=0.
\label{p-plus-minus}
\end{align}
Translated to densities and velocity fields, these become
\begin{subequations}
\begin{empheq}[box=\fbox]{align}
&\del_t \rho + \del_x(\rho v)=0,
\\
&\del_t v + \del_x(v^2/2  + \pi^2 \hbar^2 \rho^2/2 +  V(x))=0.
\end{empheq}
\label{conventional}
\end{subequations}
\kern-5pt The first equation is the continuity equation, while the second
equation is the Euler equation.
These are equations of conventional hydrodynamics.

\begin{figure}
	\begin{center}
		\begin{tabular}{ccccc}
			\begin{minipage}{0.2\hsize}
				\begin{center}
					\includegraphics[scale=.23]{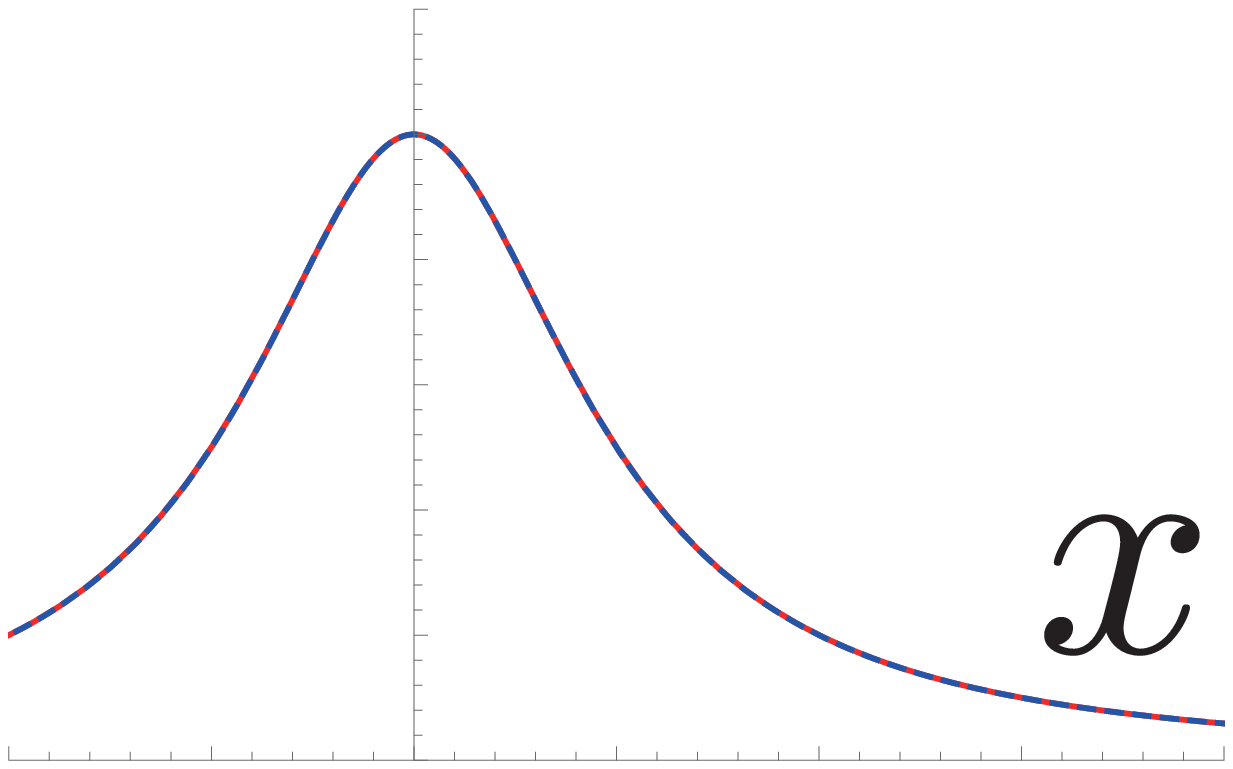}\\
					$t=0$
				\end{center}
			\end{minipage}
			\begin{minipage}{0.2\hsize}
				\begin{center}
					\includegraphics[scale=.23]{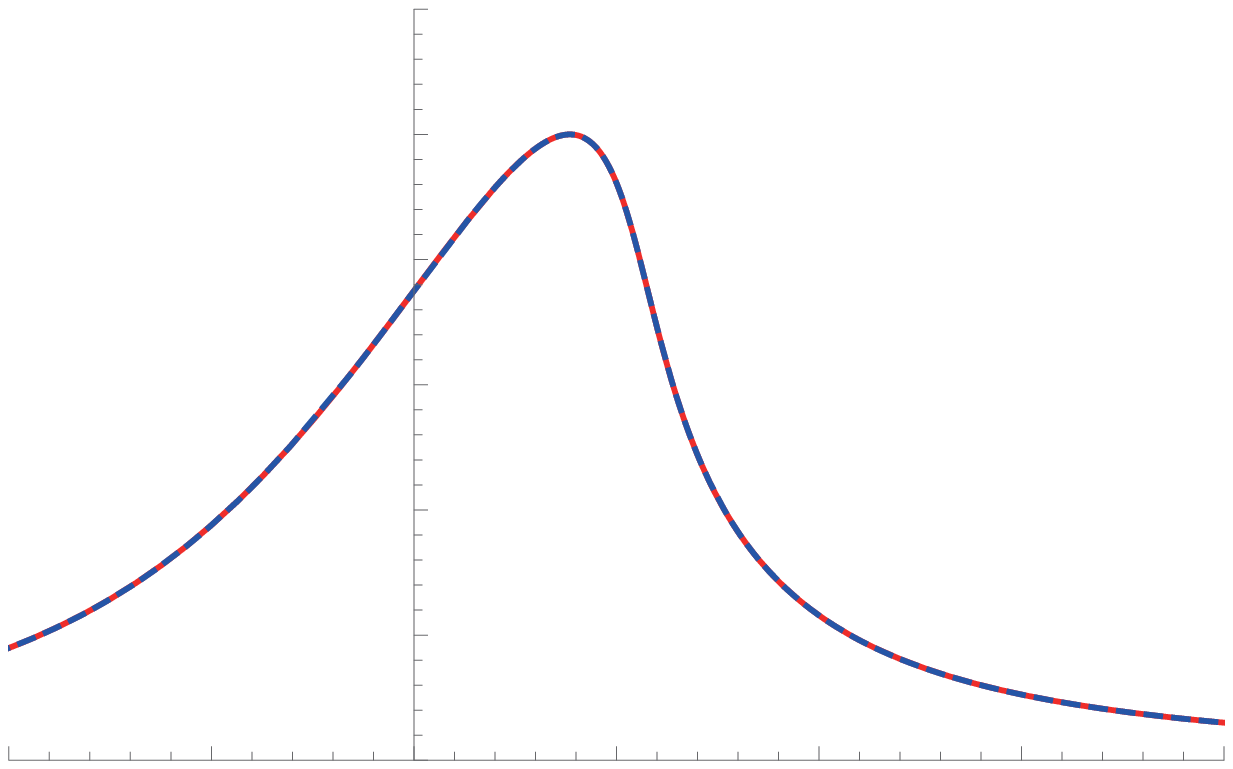}\\
					$t=0.5 t_1$
				\end{center}
			\end{minipage}
			\begin{minipage}{0.2\hsize}
				\begin{center}
					\includegraphics[scale=.23]{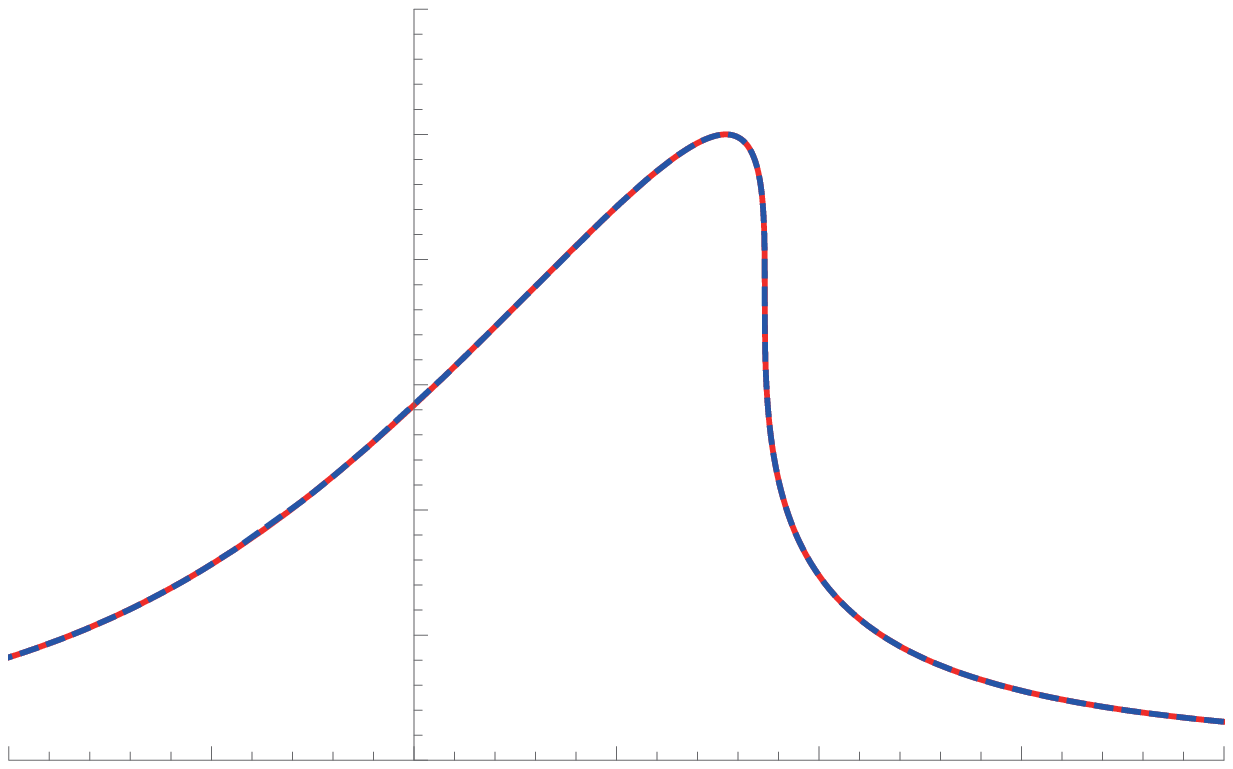}\\
					$t= t_1$
				\end{center}
			\end{minipage}
			\begin{minipage}{0.2\hsize}
				\begin{center}
					\includegraphics[scale=.23]{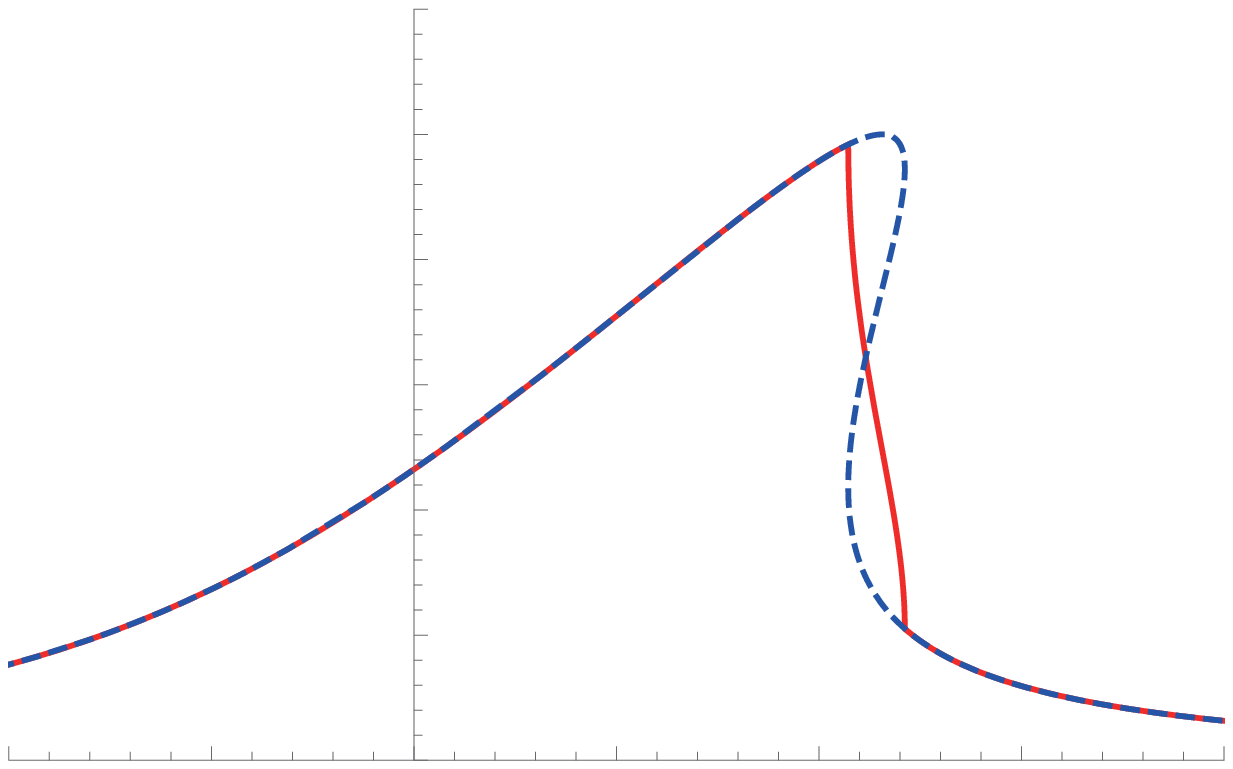}\\
					$t=1.5 t_1$
				\end{center}
			\end{minipage}
			\begin{minipage}{0.2\hsize}
				\begin{center}
					\includegraphics[scale=.23]{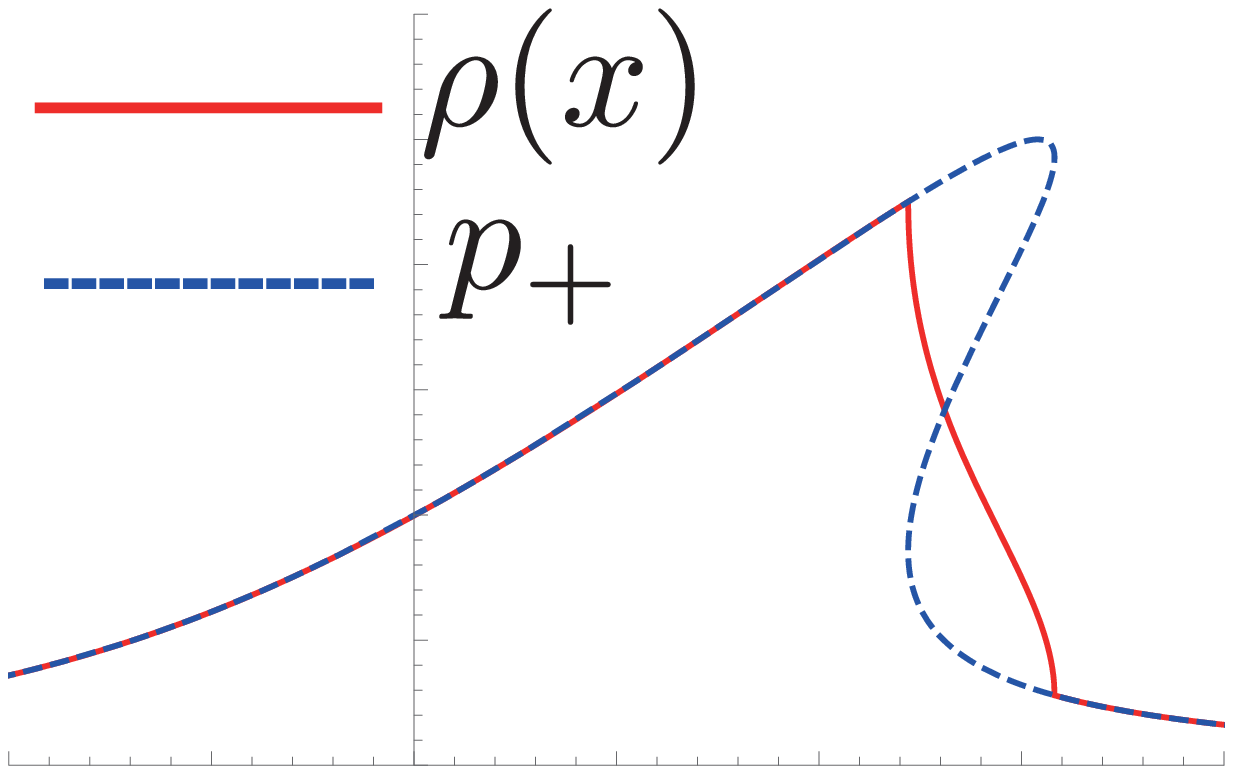}\\
					$t=2.0 t_1$
				\end{center}
			\end{minipage}
		\end{tabular}
		\caption{Development of a shock front (a single fold) in the $V=0$ case. The five panels represent
			snapshots at times $t=0, 0.5 t_1,  t_1,  1.5 t_1,  2.0 t_1$, respectively, where $t_1$ is the instant of time the overhang (fold) develops. The blue dashed curve
			represents the fluid boundary $p_+$ in the phase space, while the red curve represents the Fermion density
			$2\pi \hbar \rho(x,t)$. We have taken $p_-=0$. The horizontal axis represents $x$. 
			The $x$-turning points lead to $\del \rho/\del x = \infty$
			which characterize shock fronts. 
		}
		\label{fig-fold}
	\end{center}
\end{figure}

The Poisson bracket that leads to these equations of motion from
\eq{hamiltonian} is
\[
\{\rho(x,t), v(y,t)\}_{PB}= \del_x \delta(x-y).
\]
Now it is obvious that for a droplet of a general shape, e.g. for one with folds, the quadratic form \eq{quadratic-p} does not hold. See Figure \ref{fig-quadratic}.
Furthermore, it is easy to see that a droplet can develop folds
in time even when it does not have it initially. See Figure \ref{fig-fold}.
For a phase space
droplet which develops a fold at some time, the real space density
$\rho(x)$ develops a shock (diverging slope) at that time and the
equations of conventional hydrodynamics \eq{conventional} break
down. The phase space hydrodynamics description of Section \ref{sec:phase-hydro}, however, does not develop any pathologies since
folds are smooth configurations in phase space. 
Indeed, Figure \ref{fig-fold} is obtained via the phase space hydrodynamic calculations.

Similar analyses of shock fronts, although less extensive than the
treatment above, have
been performed in Refs.\cite{PhysRevLett.109.260602}.

{\normalsize \bibliographystyle{JHEP} \bibliography{1g} }

\end{document}